\documentclass[10pt,a4paper]{article}

\usepackage{dcolumn}
\usepackage{bm}
\usepackage{epsfig,graphicx}
\usepackage{amsmath,amsfonts,amssymb}
\usepackage{citesort}

\newcommand{\unit}{\leavevmode\hbox{\small1\kern-3.6pt\normalsize1}}

\def\lsim{\raise0.3ex\hbox{$\;<$\kern-0.75em\raise-1.1ex\hbox{$\sim\;$}}}
\def\gsim{\raise0.3ex\hbox{$\;>$\kern-0.75em\raise-1.1ex\hbox{$\sim\;$}}}
\newcommand{\wh}{\widehat}
\newcommand{\dg}[1]{#1^\dagger}
\newcommand{\vev}[1]{\langle #1\rangle}
\newcommand{\ssty}{\scriptscriptstyle}
\newcommand{\sty}{\scriptstyle}
\begin{document}

\begin{flushright}
FTUAM 05/17\\
IFT-UAM/CSIC-05-47\\
hep-ph/0512046\\
\vspace*{5mm}{\bf \today}
\end{flushright}

\begin{center}
{\Large \textbf{
FCNCs in supersymmetric multi-Higgs doublet models}}

\vspace{0.5cm} 
{\large N.~Escudero, C.~Mu\~noz and A.~M.~Teixeira} \\[0.3cm]
{\textit{Departamento de F\'{\i }sica Te\'{o}rica C-XI,\\
  Universidad Aut\'{o}noma de Madrid,
Cantoblanco, E-28049 Madrid, Spain \\[0.2cm]
Instituto de F\'{\i }sica Te\'{o}rica C-XVI,\\
  Universidad Aut\'{o}noma de Madrid,
Cantoblanco, E-28049 Madrid, Spain}}\\[0pt]

\vspace*{8mm}








\begin{abstract}
We conduct a general discussion of supersymmetric models with three
families in the Higgs sector. We analyse the scalar potential, and
investigate the minima conditions, deriving the mass matrices for the
scalar, pseudoscalar and charged states. Depending on the
Yukawa couplings and the Higgs spectrum, 
the model might allow the occurrence of potentially dangerous flavour
changing neutral currents at the tree-level. We compute model-independent  
contributions for several observables, and as an example we apply this
general analysis to a specific model of
quark-Higgs interactions,
discussing how compatibility with current
experimental data constrains the Higgs sector.
\end{abstract}
\end{center}



\section{Introduction}\label{intro}
Even though the standard model (SM)
has proved to offer a very successful description of strong and
electroweak interactions, it fails in providing an explanation to
issues such as the gauge group, the number of families, the
dynamics of flavour and the mechanism of mass generation, among others.
A particularly puzzling topic is that of the number of quark and
lepton generations. From a phenomenological and experimental 
point of view, we have 
strong reasons to accept that there are indeed only three copies of
up- and down-type quarks, as well of charged leptons and active
neutrinos. 
However, on the theoretical side, 
there is little hint to the origin
of this three-fold replication. The cancellation of 
anomalies requires identical number of quark and lepton families, but adds no
information to the total number.
Extensions of the SM, such as supersymmetry (SUSY), or grand unified
theories (GUT), remedy several deficiencies of the SM, but they also
fail in explaining the number of fermion families. 

Given the
existence of three families of quarks and leptons, and 
since neither theory nor experiment impose any constraints on the
number of Higgs families, one could wonder whether the same family
replication takes place in the Higgs sector. 
In addition to being very aesthetic, this scenario is not unexpected
from a theoretical point of view. In some SUSY models from strings, 
as is the case of the compactification of the 10 dimensional $E_8
\times E_8^\prime$ heterotic superstring to 4 dimensions, the
low-energy theory thus obtained is associated with a subgroup of
the gauge group $E_6$. Under $E_6$, fermions and
Higgses are assigned to the same representation, and the requirement of
three families of fermions imposes replication of the Higgs content
\cite{Gross:1984dd}. 
In fact, many string constructions that contain three fermion
families also include Higgs family replication, as is the case of
$Z_3$ orbifold compactifications of
the heterotic superstring, which also exhibit a low-energy spectrum composed
of three supersymmetric Higgs 
families~\cite{Casas:1988se,Casas:1988hb,Font:1989aj,Munoz:2001yj,Abel:2002ih,EMT:quark}, 
and that of some D-brane models \cite{Aldazabal:2000sa}.
Other models that include a non-minimal Higgs content have also been
extensively addressed in the 
literature~\cite{Georgi:1978ri,McWilliams:1980kj,Shanker:1981mj,Flores:1982pr,Ellis:1986ip,Cheng:1987rs,Drees:1988fc,Griest:1990vh,Griest:1989ew,Haber:1989xc,Sher:1991km,Krasnikov:1992gd,Antaramian:1992ya,Nelson:1993vc,Masip:1995sm,Masip:1995bq,Aranda:2000zf}. 
The consequences of extending the Higgs sector are abundant, and have
implications which range from the theoretical to the experimental level.
For instance, and if the extra Higgses are light, the addition of these 
states in a minimal SUSY scenario will spoil the unification of gauge 
couplings around 10$^{16}$ GeV. 
Nevertheless, in models from the heterotic string, 
the high energy scale is different 
($\approx \,g_\text{GUT}\,\times \, 10^{17}$ GeV), and the new states can even 
be helpful regarding unification~\cite{Munoz:2001yj}. Since ours is a 
low-energy oriented analysis, we will not consider the (more speculative) 
high-energy implications of the extended Higgs sector.
The most challenging implication of an extended Higgs sector is perhaps
the potential
occurrence of tree-level flavour changing neutral currents (FCNC),
mediated by the exchange of neutral Higgs states\footnote{
For other consequences such as flavour-violating Higgs and top 
decays, and the associated
experimental signatures at the next generation of colliders,
see, for example,~\cite{Bejar:2005kv,Curiel:2003uk,Aguilar-Saavedra:2004wm}  
and references therein.}. 
In the SM and the minimal supersymmetric standard model (MSSM), 
these effects are absent at the tree-level,
since the coupling of the quark-quark-Higgs mass eigenstates is flavour
conserving. This arises from having the Yukawa couplings proportional
to the quark mass matrices, so that 
diagonalising the mass matrices also diagonalises the Yukawas.
Since experimental data is in good agreement with the SM predictions,
the potentially large contributions arising from the tree-level
interactions must be suppressed in order to have a model which is
experimentally viable. In general, the most stringent limit on the
flavour-changing processes emerges from the small value of the
$K_L-K_S$ mass difference~\cite{McWilliams:1980kj,Shanker:1981mj}.

Avoiding the FCNCs induced by the tree-level exchange of neutral Higgs 
bosons can be achieved by several distinct approaches, each involving a
different sector of the model. 

\noindent 
(i) Discrete symmetries. Imposing a discrete symmetry on the model
ensures that only one generation of Higgs couples to quark and
leptons, and completely eliminates tree level FCNCs from the
predictions of the model~\cite{Glashow:1976nt,Paschos:1976ay}. 
Naturally, one attempts to motivate such an assumption via
topological and/or geometrical arguments.
For instance, in~\cite{Griest:1989ew,Griest:1990vh}, adding
four extra Higgs doublets to the MSSM content, and assuming that the
FCNCs thus induced are suppressed via some symmetry, then the extra
generations (labelled ``pseudo-Higgs bosons'') do not acquire 
vacuum expectation values (VEVs), do not 
mix with the two MSSM-like doublets, nor couple to fermions. Still, the
lightest of the new states is stable so that it becomes a candidate
for dark matter. 

\noindent 
(ii) Suppression of Yukawa couplings. In this case, FCNCs are not
eliminated, but the Yukawa interaction responsible for the tree-level
flavour violation is made small enough to render the new contributions
negligible. Since the most stringent bounds are usually associated
with tree level contributions to the neutral kaon mass difference ($\Delta
m_K$), the Yukawa couplings of down and strange quarks are forced to
be very small. 
For example, in \cite{Aranda:2000zf}, a $U(2)$ flavour symmetry was
considered as the candidate symmetry to suppress FCNCs.
Although this second possibility is more appealing in the sense that
it does not require excluding part of the Higgs content from
Higgs-matter interactions, its major shortcoming lies in the fact that
in general one lacks a full theory of flavour, which would predict the
Yukawa couplings\footnote{Nevertheless, it is worth noticing that
 in the framework of the MSSM, there are
 several models where it is found that via flavour symmetries one
 can successfully explain the observed pattern of masses and
 mixings.}, as would be the case of string theory.

\noindent 
(iii) Decoupling of extra Higgses. If one does not wish (or is not
allowed) to impose a symmetry, or if the Yukawa couplings are not free
parameters of the model, a third possibility lies in making the new
Higgs states heavy enough, so that the contributions they mediate are
suppressed~\cite{Georgi:1978ri,Cheng:1987rs,McWilliams:1980kj,Shanker:1981mj,Sher:1991km}. 
The new states are thus decoupled, and the effective 
low-energy theory is very similar to the usual MSSM/SM. 
This possibility is sometimes the only ``degree of freedom''
remaining, especially in highly predictive models, where the Yukawas
arise from some high-energy formulation, as is the case of string models.
Nevertheless, it is important to stress that the decoupling scenario
is in general achieved by enforcing very large values for some of the Higgs
soft-breaking masses. This may lead to a fine-tuning scenario in
association with electroweak (EW) symmetry breaking.
When the decoupling approach is taken together with the suppression of 
Yukawa couplings (ii), it is
possible to obtain Higgs spectra which manage to comply with
experiment without excessively heavy Higgses~\cite{McWilliams:1980kj,Shanker:1981mj,Cheng:1987rs,Sher:1991km,Krasnikov:1992gd,Antaramian:1992ya,Aranda:2000zf}.

From the above discussion, it is clear that a correct evaluation of
the FCNC problem associated to multi-Higgs doublet models is indeed
instrumental. In what follows, we propose to analyse the most general 
scenario of the MSSM extended to include three families of Higgs
doublets. We do
not impose any symmetry on the model, allowing for the most general
formulation of both the superpotential and the SUSY soft breaking
Lagrangian. We provide a generic overview of the extended Higgs
sector, and discuss the minimisation of the scalar potential. 
In this work, we also compute the most general expression for the
contribution of tree-level 
neutral Higgs exchange to neutral meson mass difference. Contrary to
previous analyses~\cite{McWilliams:1980kj,Shanker:1981mj,Cheng:1987rs,Sher:1991km}, we include the
contributions from all physical (rather than interaction) Higgs states,
scalar and pseudoscalar. Moreover, we take into account the mixing in
the Higgs sector.
Finally, as an example of how to apply our general formulation, we
consider an ansatz for the Yukawa couplings along
the lines of previous analyses (namely the ``simple Fritzsch 
scheme''~\cite{Fritzsch:1977vd,Fritzsch:1999ee,Sze:2005gc}), and
evaluate the specific contributions to the neutral mesons mass
differences. Our findings turn out to be more severe than those of previous
works. 

The work is organised as follows.
In Section~\ref{higgsphenom}, we analyse the extended
Higgs sector, paying special attention to the minimisation of the
Higgs potential and addressing potential fine-tuning issues. 
We compute the tree level mass matrices in Section~\ref{higgsspectrum} 
and discuss the associated spectra.
Section~\ref{yukint} is devoted to the analysis of Higgs-matter
interactions, and in Section~\ref{fcnc} we derive a model independent 
computation of the
tree-level contributions to the neutral meson mass differences. 
Assuming an illustrative example for the Yukawa couplings, we present
in Section~\ref{flavour:res} a numerical study of the FCNC contributions,
investigating how the Higgs spectrum must be constrained in order to
have compatibility with experimental data. Finally we summarise the
most relevant points in Section~\ref{conc}.

\section{Extended Higgs sector}\label{higgsphenom}
We begin our analysis by addressing the Higgs sector of a SUSY model where
three generations of $SU(2)$ doublet superfields are comprised.
In each generation, one finds hypercharge $-1/2$ and $+1/2$ fields, 
coupling to down- and up-quarks, respectively:
\begin{equation}\label{H:superf}
\widehat{H}_{1(3,5)}=
\left( \begin{array}{c}
\widehat{h}^0_{1(3,5)} \\
\widehat{h}^-_{1(3,5)}
\end{array} \right)\,,  \quad \quad
\widehat{H}_{2(4,6)}=
\left( \begin{array}{c}
\widehat{h}^+_{2(4,6)} \\
\widehat{h}^0_{2(4,6)}
\end{array} \right)\,.
\end{equation} 

\subsection{Tree-level Higgs potential}
The most general superpotential of a model with three families of 
Higgs doublets can be written as follows:
\begin{align}
W &=\,
\wh{Q}\,
(Y_1^d\wh{H}_1+Y_3^d\wh{H}_3+Y_5^d\wh{H}_5)\wh{D}^c+
\wh{L}\,(Y_1^e\wh{H}_1+Y_3^e\wh{H}_3+Y_5^e\wh{H}_5)\wh{E}^c
 \nonumber \\ 
& +\, \wh{Q}\,
(Y_2^u\wh{H}_2+Y_4^u\wh{H}_4+Y_6^u\wh{H}_6)\wh{U}^c+\mu_{12}\wh{H}_1\wh{H}_2+
\mu_{14}\wh{H}_1\wh{H}_4+\mu_{16}\wh{H}_1\wh{H}_6 \nonumber
\\ & + \,\mu_{32}\wh{H}_3\wh{H}_2+\mu_{34}\wh{H}_3\wh{H}_4+
\mu_{36}\wh{H}_3\wh{H}_6+\mu_{52}\wh{H}_5\wh{H}_2+\mu_{54}\wh{H}_5\wh{H}_4+
\mu_{56}\wh{H}_5\wh{H}_6\,.
\end{align}
Above, $\wh{Q}$ and $\wh{L}$ denote the quark and lepton $SU(2)_L$ 
doublet superfields, 
$\wh{U}^c$ and $\wh{D}^c$ are quark singlets, $\wh{E}^c$ the lepton
singlet, and $Y_i^q$ are the Yukawa matrices associated with each
Higgs superfield. Regarding the $\mu$-terms, these are now extended in
order to include all possible bilinear Higgs terms. Calling upon some
specific model that would be responsible for generating the latter
terms, it would be possible to reduce the number of free
parameters. It is also important to recall that one could also impose a
(discrete) symmetry acting on the superpotential, whose effect
would be the suppression of some of the couplings in $W$.
However, in our discussion, we consider the most generic form for $W$,
thus taking the several bilinear parameters in $W$ as effective $\mu$-terms.

From the above one can derive the $F$- and $D$-terms, 
writing the latter in doublet component for simplicity:
\begin{align}
V_F\,
=&
\operatornamewithlimits{\sum}_{\begin{smallmatrix}
{i,j=1,3,5}\\{l=2,4,6}
\end{smallmatrix}} \mu^*_{il} \,\mu_{jl}\, H_i^\dagger\, H_j
+
\operatornamewithlimits{\sum}_{\begin{smallmatrix}
{i=1,3,5}\\{k,l=2,4,6}
\end{smallmatrix}} \mu^*_{il} \,\mu_{ik}\, H_k^\dagger\, H_l \,,\nonumber\\
V_D\,
=&\,\,
\frac{g^2}{8} \,\operatornamewithlimits{\sum}_{a=1}^{3} 
\left[ \operatornamewithlimits{\sum}_{i=1}^6
H_i^\dagger \,\tau^a \,H_i
\right]^2 + \,
\frac{g^{\prime 2}}{8} 
\left[\operatornamewithlimits{\sum}_{i=1}^6 
(-1)^i \left|H_i\right|^2 \right]^2\,,
\end{align}
where $H_i$ are the scalar doublets belonging to the
superfields of Eq.~(\ref{H:superf}), 
$g$, $g^\prime$ denote the gauge coupling constants, and
$\tau^a$ are the $SU(2)_L$ generators.
The soft SUSY-breaking terms trivially generalise the minimal
two-doublet case, including the usual soft-breaking masses 
and $B\mu$-like terms in the Higgs potential. 
\begin{align}
V_{\text{soft}}=&
\operatornamewithlimits{\sum}_{i,j=1,3,5}
(m^2_d)_{ij} \, H_i^\dagger \, H_j
+
\operatornamewithlimits{\sum}_{k,l=2,4,6}
(m^2_u)_{kl} \, H_k^\dagger \, H_l
-\operatornamewithlimits{\sum}_{\begin{smallmatrix}
{i=1,3,5}\\{l=2,4,6}
\end{smallmatrix}}
\left[(B\mu)_{il}\, H_i\, H_l +\text{H.c.}\right]\,.
\end{align}

\subsection{Minimisation of the potential: the
  superpotential basis and the Higgs basis}
After electroweak symmetry breaking, the neutral components of the six
Higgs doublets develop the following VEVs, 
\begin{equation}
\vev{h^0_{1(3,5)}}=w_{1(3,5)}\,, 
\quad \quad \quad 
\vev{h^0_{2(4,6)}}=w_{2(4,6)}\,,
\end{equation}
and as usual one can write
\begin{align}
h^0_i \to w_i + \frac{1}{\sqrt{2}} \left(\sigma_i + i \varphi_i\right)\,.
\end{align}
In the above we assume that all the VEVs are real\footnote{We do not
discuss the possibility of spontaneous CP violation in association
with this class of multi-Higgs doublet models.}.
The next step is to minimise the scalar potential with respect to the
VEVs. 
Although the problem appears to be a simple 
generalisation of the two-Higgs doublet model (MSSM), it will prove to
be much more involved.
It happens that in the presence of six non-vanishing VEVs, finding a
solution to the minima equations is a rather cumbersome task. 
As an illustration, we
present the equations obtained from minimising with respect to the
down-type Higgs:
\begin{align}\label{min:downs}
\sum_{i=1,3,5} \left[
(m^2_d)_{ji} +
\sum_{l=2,4,6} \mu_{jl}^* \, \mu_{il}\, \right] 
w_i 
-\sum_{l=2,4,6} (B\mu)_{jl} \,w_l + w_j\, \frac{1}{2}\, M_Z^2\,
\frac{v_d^2-v_u^2}{v_d^2+v_u^2} =0\,,
\end{align}
where $j=1, 3, 5$ and we have defined
\begin{align}\label{vuvd:def}
v_d \equiv & \sqrt{w_1^2+w_3^2+w_5^2} \,,\nonumber \\
v_u \equiv & \sqrt{w_2^2+w_4^2+w_6^2}\,.
\end{align}
In addition to Eq.~(\ref{min:downs}), which only ensure that we are in
the presence of an extremum, we have to further impose the conditions
for a minimum 
with respect to the several variables. 
Therefore, and for the down-type Higgs fields, having positive second
derivatives is equivalent to the following inequalities:
\begin{align}\label{min:downs:2}
(m^2_d)_{jj} +
\sum_{l=2,4,6} \mu_{jl}^* \,\mu_{jl} + 
\frac{1}{2}\, M_Z^2 \, \left(\,
\frac{v_d^2-v_u^2+2 \, w_j^2}{v_d^2+v_u^2}\,\right)
\,>\,0 \,,
\end{align}
provided that at the minimum the determinant of the
six-dimensional matrix is positive, or equivalently, $\det (\partial^2\, V\,/\,
\partial h^0_i\,\partial h^0_j)\,>0$. 

The equations for the remaining cases (minimising
with respect to the up-type Higgs fields) can be obtained from
Eqs.~(\ref{min:downs},\ref{min:downs:2}) by interchanging 
$\{1,3,5\} \leftrightarrow \{2,4,6\}$ and changing the sign in the
terms proportional to $M_Z^2$ as $M_Z^2 (v_d^2-v_u^2)
\to - M_Z^2 (v_d^2-v_u^2)$ (cf. Eq.~(\ref{min:downs})) and 
$M_Z^2 (v_d^2-v_u^2+2 w^2)
\to - M_Z^2 (v_d^2-v_u^2-2 w^2)$ (cf. Eq.~(\ref{min:downs:2})).
The only choice of parameters that would render the solution of the
minima equations straightforward is to call upon the diagonal soft
breaking masses ($(m_d)_{ii}$ and $(m_u)_{kk}$) to be fixed by the
minima equations. For $(m_d)^2_{11}$ the minimum equation would read
\begin{align}\label{md11:min}
(m_d)^2_{11} =& \sum_{j=2,4,6} (B\mu)_{1j}  \,\frac{w_j}{w_1} -
(m^2_d)_{13}  \,\frac{w_3}{w_1} - (m^2_d)_{15}  \,\frac{w_5}{w_1} \nonumber \\
&- \sum_{i=1,3,5}  \,( \sum_{l=2,4,6} \mu_{1l}^* \mu_{il}  \,
\frac{w_i}{w_1}  \,) - \frac{1}{2} M_Z^2 \,
\frac{v_d^2-v_u^2}{v_d^2+v_u^2}  \,.
\end{align}
Naturally, the equations can be solved for a distinct
set of parameters, but then finding an analytical solution is in
general not possible. Analytical relations are very useful when
writing down the mass matrices for the several Higgs sectors, since
one can directly replace the parameters. In fact, the major handicap
when relying on the diagonal soft masses as the minimum parameters is
that then one loses control of the leading contributions to the Higgs
boson mass eigenvalues. This will be made clear in a forthcoming
section. Another shortcoming is that, even though we can define a
generalised expression for $\tan \beta$ as 
$\tan \beta= \sqrt{w_2^2+w_4^2+w_6^2}\, /\,\sqrt{w_1^2+w_3^2+w_5^2}$, 
writing the minimisation equations as a function of $\sin \beta$,
$\cos \beta$ (in an MSSM-like fashion), is impossible.

So far we have been working in the basis that naturally emerges
from both the superpotential and soft-breaking Lagrangian
formulation. To simplify the discussion, we will
henceforth label this natural basis as the superpotential basis.
However, this need not be the unique approach when addressing the
minimisation of the potential.

As it has been shown, one can work in a
basis where only two of the new six neutral fields have non-vanishing
VEVs, the so-called Higgs basis~\cite{Georgi:1978ri,Drees:1988fc}. For the
down-type Higgses, the new fields, $\phi_i$, are related to the 
original ones by means of the following unitary transformation, $P_d$:
\begin{align}
& \phi^0_1 \equiv 
\frac{1}{\sqrt{w_1^2+w_3^2+w_5^2}}\,
\left(w_1 \,h^0_1+w_3 \,h^0_3+w_5 \,h^0_5\right)\,, \\
& \phi^0_3 \equiv 
\frac{1}{\sqrt{w_1^2+w_3^2+w_5^2}}
\left[\sqrt{w_3^2+w_5^2}\, h^0_1-
\frac{w_1}{\sqrt{w_3^2+w_5^2}}(w_3 \,h^0_3+w_5 \,h^0_5)\right]\,, \\
& \phi^0_5 \equiv
\frac{1}{\sqrt{w_3^2+w_5^2}}\left(w_5 \,h^0_3-w_3 \,h^0_5\right)\,.
\end{align}
An analogous transformation, $P_u$, can be derived for 
the up-quark-coupling Higgses ($h^0_{2,4,6}$), with
the adequate replacements ($\{ 1,3,5\} \to \{ 2,4,6\}$). 
For simplicity, let us introduce the following global parametrisation for
the transformations $P_d$ and $P_u$:
\begin{align}\label{higgs:Ptransf}
& \phi_i = P_{ij} h_j\,,
\end{align}
where $P_{ij}$ is a $6 \times 6$ matrix whose entries are defined from
$P_d$ and $P_u$ as
\begin{align}\label{higgs:Ptransf:2}
& P_{ij}= \left\{
\begin{array}{ll}
(P_d)_{\frac{i+1}{2} \frac{j+1}{2}} &\,\, \,i,j=1,3,5\,\\
(P_u)_{\frac{i}{2} \frac{j}{2}} & \,\, \,i,j=2,4,6 \,\\
\,\, \,\,0 & \,\, \, \,\,\text{elsewhere}\,.
\end{array}\right.
\end{align}
From the above, it is clear that in the new basis only two of the
fields do have a VEV
\begin{equation}
\vev{\phi^0_1} =\,  v_d\,, \quad \quad
\vev{\phi^0_2} =\,  v_u\,,
\end{equation}
with $v_{d,u}$ defined in Eq.~(\ref{vuvd:def}), which in turn must satisfy 
\begin{equation}\label{ewz}
  v_u^2+v_d^2=2\,M_Z^2/(g^2+g'^2)\approx (174\text{ GeV})^2\,,
\end{equation}
where $M_Z$ is the mass of the $Z^0$ boson.
In this basis one has the additional advantage that, similar to what
occurs in the MSSM case, one can clearly define $\tan \beta$:
\begin{equation}\label{tb}
\tan \beta = \frac{v_u}{v_d}\,.
\end{equation}
We can now write
\begin{align}\label{higgsbasis:decomp}
&\phi^0_{1,2} \to v_{d,u} +\frac{1}{\sqrt{2}} \left(
R_{1,2} + i I_{1,2}\right)\,, \nonumber\\
&\phi^0_{i} \to \frac{1}{\sqrt{2}} \left(R_{i} + i I_{i}\right)\, 
\quad \quad i=3-6\,.
\end{align}
Thus, one can now understand the three Higgs family model as an MSSM-like
model, extended by 4 additional doublets, which do not directly interfere with
the mechanism of electroweak symmetry breaking. 
In the Higgs basis, the potential preserves its original structure
with respect to the dependence on the field combination,
but the parameters associated with the $F$- and soft breaking terms are now
redefined as 
\begin{align}
V_F \,
=&
\operatornamewithlimits{\sum}_{\begin{smallmatrix}
{a,b=1,3,5}\\{l=2,4,6}
\end{smallmatrix}} \bar\mu^l_{ab}  \,\phi_a^\dagger \, \phi_b
+
\operatornamewithlimits{\sum}_{\begin{smallmatrix}
{i=1,3,5}\\{c,d=2,4,6}
\end{smallmatrix}} \bar{\bar{\mu}}^i_{cd} \, \phi_c^\dagger \, \phi_d \,,
\nonumber\\
V_{\text{soft}} \,
=&
\operatornamewithlimits{\sum}_{a,b=1,3,5}
(\bar m^2_d)_{ab}  \, \phi_a^\dagger  \,\phi_b
+
\operatornamewithlimits{\sum}_{c,d=2,4,6}
({\bar{\bar m}}^2_u)_{cd}  \, \phi_c^\dagger \, \phi_d
-\operatornamewithlimits{\sum}_{\begin{smallmatrix}
{a=1,3,5}\\{c=2,4,6}
\end{smallmatrix}}
\left[(\overline{B\mu})_{ac}  \,\phi_a  \,\phi_c +\text{H.c.}\right]\,,
\end{align}
where
\begin{align}
&\bar\mu^l_{ab}=(P_d)_{ai}  \,\mu^*_{il} \, \mu_{jl} \, (P_d^\dagger)_{jb}\,,
\quad \quad \quad \quad \quad
\bar{\bar{\mu}}^i_{cd}=(P_u)_{cl} \, \mu^*_{il}  \,\mu_{ik}  \,
(P_u^\dagger)_{kd}\,,
\nonumber \\
&(\bar m^2_d)_{ab} = (P_d)_{ai} \, (m^2_d)_{ij}  \,(P_d^\dagger)_{jb}\,,
\quad \quad \quad
({\bar{\bar m}}^2_u)_{cd}=(P_u)_{ck}  \,(m^2_u)_{kl} \, (P_u^\dagger)_{ld}\,,
\nonumber \\
&(\overline{B\mu})_{ac}= (P_d^\dagger)_{ia}  \,(B\mu)_{il} \,
(P_u^\dagger)_{lc}\,,
\quad \quad \quad i,j,a,b=1,3,5\,; \quad \quad
k,l,c,d=2,4,6\,.
\end{align}
In the new basis, 
the problem of minimising the scalar potential is quite simplified,
and one is able to derive the six minimisation conditions in a compact
form. The conditions for the down-type sector 
read
\begin{align}\label{minima:d}
m^2_{11}\,=&\, \,
b_{12} \,\tan \beta - \frac{M_Z^2}{2}\,\cos 2 \beta\,, 
\nonumber \\
m^2_{13}\,=&\, \,
b_{32} \,\tan \beta \,,\nonumber \\
m^2_{15}\,=&\, \,
b_{52} \, \tan \beta \,,
\end{align}
while those related to the up-type Higgses are given by
\begin{align}\label{minima:u}
m^2_{22}\,=&\, \,
b_{12} \,\cot \beta + \frac{M_Z^2}{2}\,\cos 2 \beta\,, 
\nonumber \\
m^2_{24}\,=&\, \,
b_{14}\, \cot \beta\,, \nonumber \\
m^2_{26}\,=&\, \,
b_{16} \,\cot \beta \,.
\end{align}
In the above, and for simplicity, we have introduced the short-hand notation
\begin{align}\label{mij:bij:def}
m_{ij}^2 &= \left\{
\begin{array}{l}
\operatornamewithlimits{\sum}_{l=2,4,6}
 \,\bar\mu^{l}_{ij}+(\bar{m}_d^2)_{ij} \quad \quad i,j=1,3,5
\nonumber \\
\operatornamewithlimits{\sum}_{k=1,3,5}
 \,\bar{\bar\mu}^{k}_{ij}+(\bar{\bar{m}}_u^2)_{ij} \quad \quad 
i,j=2,4,6 \,,
\end{array}
\right. \\
b_{ij} &=(\overline{B\mu})_{ij}\,,
\end{align}
where we stress that both parameters $\bar\mu_{ij}$, and $\bar{\bar\mu}_{ij}$,
as well as $b_{ij}$ have dimensions (mass$^2$).
Throughout the paper we will be often using the above short-hand
notation, even though this implies that the rotated $\mu$-terms and
Higgs soft breaking masses loose their individual character, becoming
merged into the quantities $m^2_{ij}$.
As a final comment, let us notice that
the minima equations (\ref{minima:d},\ref{minima:u}) 
are, in structure, very similar to those of the
MSSM, namely the first in each set of three. It is worth referring
that the EW scale is only explicitly 
present in two equations (those derived with
respect to $\phi^0_{1,2}$, the VEV-acquiring fields).
Notice however, that this is a mere consequence of working in a
different basis. 
 
In addition to the minima conditions
Eqs.~(\ref{minima:d},\ref{minima:u}), one can also derive other
conditions, which are useful in constraining the allowed parameter
space. As an example, and from direct comparison of the minima
equations for $m^2_{11}$ and $m^2_{22}$, we find the following
inequality:
\begin{equation}
m^2_{11} \, m^2_{22}\, \leq\, b^2_{12}\,,
\end{equation}
 strongly resembling the analogous MSSM condition. 

\subsection{Numerical minimisation of the potential: illustrative
  examples}\label{numv}
In what follows, our goal is to conduct a short analysis of
the issues discussed in the previous subsections.
The phenomenological analysis of the Higgs sector 
(masses and mixings) will be carried in detail in 
Section~\ref{higgsspectrum}, working in the Higgs basis. 
Nevertheless, we find it interesting to evaluate
how the minimisation of the scalar potential constrains the soft breaking
terms. To do so, we
consider several numerical examples for the minimisation of the
scalar potential for distinct VEV regimes. In addition, we also discuss 
whether or not a specific choice of soft breaking terms may
translate into a fine-tuning problem. Although the minima conditions
are cast in a much more compact and appealing way in the Higgs
basis, since only two of them explicitly involve the EW scale, a
possible fine tuning problem may be unapparent in this
case. Therefore, for this specific discussion, we will stick to the 
original superpotential basis.
In this case, and since we make no assumptions regarding the mechanism of SUSY
breaking associated with this generic multi-Higgs doublet model,
we take the input parameters 
\begin{equation}\label{min:free:para}
(m_d^2)_{ij}\,, \ \ (m_u^2)_{ij}\,, \ \ \mu_{ij}\,, \ \ (B \mu)_{ij}\,,
\end{equation}
to be free at the electroweak scale. Regarding the Higgs VEVs ($w_i$), we
impose no constraint other than satisfying the EW
breaking conditions for a given value of $\tan \beta$. 
In order to further simplify the approach, we
will fix, in each separate analysis, the value of $\mu_{ij}=\mu$, and 
assume a common overall
scale for the off-diagonal masses $(m_d)_{ij}$ and $(m_u)_{ij}$, and
for $(B \mu)_{ij}$. More specifically, we take universal $B\mu$ terms, 
$(B \mu)_{ij}= B\mu = M_S^2$, where $M_S$ 
is a ``typical soft-SUSY breaking scale''. For the 
off-diagonal masses $(m_d)_{ij}$ and $(m_u)_{ij}$, the common scale is
taken to be $(m_d)_{ij}=(m_u)_{ij}=0.1\, M_S$\footnote{
Notice that for $i \neq j$, values of $(m_q)_{ij} \gtrsim M_S$ are not
compatible with viable solutions. This is evident from the minimisation
conditions of Eq.~(\ref{md11:min}), using $(m_q)^2_{ij} >0$.
Furthermore, the limit $(m_q)_{ij} \to 0$ occurs in well
motivated models~\cite{Brignole:1997dp}.}.

As discussed in the previous subsection, 
the parameters which are (trivially) fixed by the six minima equations
are the diagonal Higgs soft-breaking masses, $(m_d^2)_{ii}$ and 
$(m_u^2)_{ii}$. Clearly, from
Eq.~(\ref{md11:min}), the minima values for $(m^2_q)_{ii}$, and in
particular those of the down-type Higgs sector, will be very dependent
of the values of the VEVs considered, most specifically on whether one
chooses a degenerate or hierarchical regime. Therefore, each 
case has to be separately investigated, and in each
situation, one must analyse the impact of the other free parameters,
namely that of the overall scale chosen for the soft-breaking parameters.

In addition, one has to investigate whether or not the parameters taken
lead to a potential fine tuning (FT) problem for the
parameters. In doing so, our approach will be the following. For a
specific scenario defined by the parameters in
Eq.~(\ref{min:free:para}), together with $w_i$ and $\tan \beta$, we
compute the values of the diagonal soft breaking masses, as imposed
from complying with the minima equations. We then impose a tiny
perturbation on the 
solutions,
\begin{equation}\label{Dmq:def}
(m^2_q)_{ii}\, \to\, 
(m^2_q)_{ii}^\prime \,= \,(1+\lambda^q)\, (m^2_q)_{ii}\,,
\end{equation}
where $\lambda$ is taken to be around a few percent.
Finally, we compute the new value of $M_Z$ derived from perturbing the
correct (true minima) solution for $(m^2_q)_{ii}$, $M_Z^\prime$.
To evaluate the amount of FT, we follow~\cite{Barbieri:1987fn,Casas:2004gh}, 
and introduce the parameter $\Delta p$, defined as 
\begin{equation}\label{Dp:def}
\Delta p_i \, \frac{\delta p_i}{p_i} \,=\,
 \frac{\delta M_Z^2}{M_Z^2}\,,
\end{equation}
identifying $p_i$ with $(m^2_q)_{ii}$. A rough measure of the
fine-tuning can be derived from $(\Delta p_i)^{-1}$, since the latter
can be identified with the probability of a cancellation between terms
of a given size to obtain a final result $\Delta p_i$ times smaller.

In what follows we take, as illustrative examples, two
distinct cases for the Higgs VEVs: degenerate and hierarchical
$w_i$. We stress here that at this point we are only investigating the
constraints on the soft breaking terms arising from the minima
conditions. A study of the spectrum is far simplified
when moving to Higgs basis, as will be done in the following Section.\\

\noindent
$\bullet$ Degenerate VEVs: $w_1=w_3=w_5$, $w_2=w_4=w_6$

\noindent
In this case, the minima equations become much simpler, since in each
sector the VEVs factor out.
Therefore, having fixed the free parameters as above described,
the six original
equations essentially reduce to two independent ones: one for the
down-type Higgs (e.g. $(m_d)^2_{11}$) and another for the up-type
sector (e.g. $(m_u)^2_{22}$).

\begin{figure}[t] 
\begin{center}
    \begin{tabular}{cc}
\psfig{file=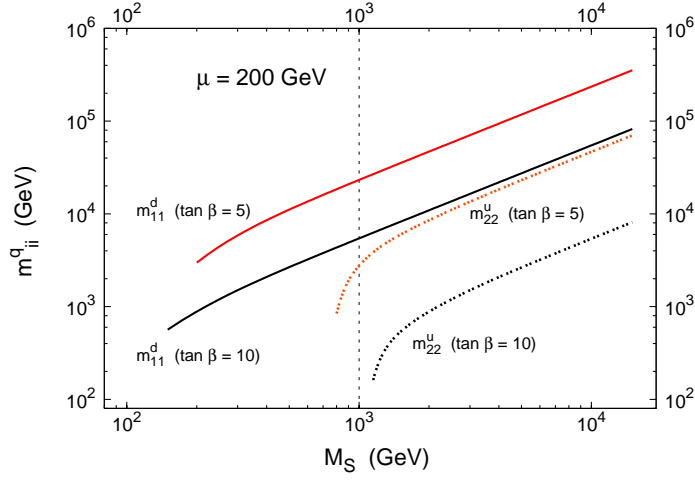,angle=270,width=100mm,clip=}
    \end{tabular}
    \caption{Minima solutions $m^d_{11}$ and $m^u_{22}$ as a function
    of the common SUSY breaking scale $M_S$ 
    for $\tan \beta=$5, 10 and degenerate VEVs. 
    To the left of the vertical dotted line, the solutions for $m^q_{ii}$
    are not true minima of the potential.}
    \label{fig:mh2out.msusy.k1}
  \end{center}
\end{figure}

In Fig.~\ref{fig:mh2out.msusy.k1}, 
we plot $(m_d)^2_{11}$ and $(m_u)^2_{22}$, as computed
from the tree-level minima equations, as a function of the common soft
breaking term scale, $M_S $, for $\mu_{ij}=\mu=200$
GeV. In each case, we consider two distinct values for $\tan
\beta$: 5 and 10.
From Fig.~\ref{fig:mh2out.msusy.k1}, several properties of this model
become apparent. 
First of all, it becomes clear that in this basis (the superpotential
basis) finding minima can be a challenging task. Notice that although
one can find solutions for the down-type sector associated with low 
$M_S$, in this range we are in the presence of maxima (rather than
minima) for the up-type soft masses. The situation slightly worsens
with increasing $\tan \beta$. Taking other (higher) values of $\mu$
would lead to the displacement of the vertical line further to the right of the
parameter space parametrised by $M_S$.
This behaviour can be confirmed by inspection of 
Eqs.~(\ref{min:downs},\ref{min:downs:2},\ref{md11:min}).

\begin{figure}[t]
\begin{center}
\psfig{file=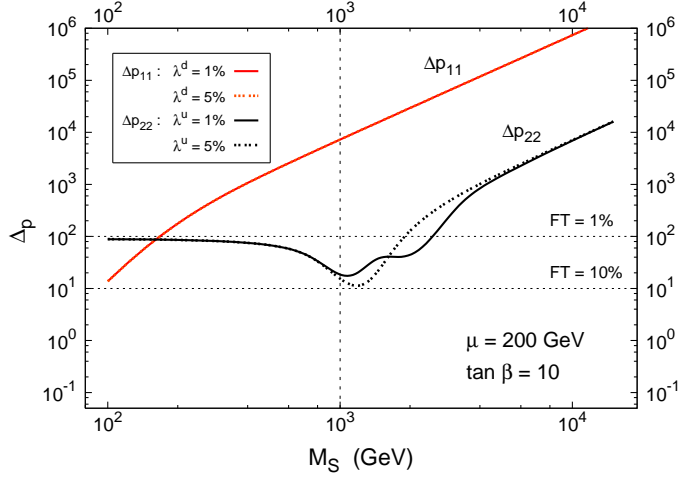,angle=270,width=100mm,clip=}
    \caption{$\Delta p$ as a function of the 
      SUSY scale $M_S$ for perturbations of 1\% and 5\% 
      in $(m^d_{11})^2$ and $(m^u_{22})^2$. 
      $\mu=200$ GeV, and  $\tan \beta=10$ (the two lines for $\Delta
      p_{11}$ appear over-set). 
      To the left of the vertical dotted
    line, the solutions for $m^q_{ii}$ are not true minima of the potential.
    Horizontal lines denote areas above which
    one roughly expects a FT stronger than 1\% and 10\%.}
    \label{fig:Dpout.msusy.k1}
  \end{center}
\end{figure}

Let us now evaluate how fine-tuned these solutions are. 
In Fig.~\ref{fig:Dpout.msusy.k1}
we investigate $\Delta p$ for the up and down sectors, by studying in
each case the effect of perturbing the minima solutions $(m^d_{11})^2$
and $(m^u_{22})^2$ by $\lambda^{d,u}= 1\%$ and 
$\lambda^{d,u}= 5\%$. We again consider $\mu=200$ GeV, and take $\tan
\beta=10$.  
 
From Eqs.~(\ref{Dmq:def},\ref{Dp:def}) together with the minimisation
conditions in Eq.~(\ref{md11:min}), it is apparent that 
$\Delta p_i \propto \frac{(m_q^2)_{ii}}{M_Z^2}$.
Moreover, it can be clearly seen that for the down-type sector, the
dominant term ($B \mu$) is further enhanced by a factor of $\tan
\beta$, so that a tiny fluctuation in $(m^d_{11})^2$ is not easily
cancelled. In the up-type masses, the situation is reversed. The
up-type version of  Eq.~(\ref{md11:min}) would exhibit a suppression
of $B \mu$ by $\cot \beta$, so that here we find a more relaxed scenario.\\

\noindent
$\bullet$ Hierarchical VEVs: $w_5 = 10\,w_3=100\,w_1$, 
$w_6= 10\,w_4=100\,w_2$ 

\begin{figure}[t] 
\begin{center}
    \begin{tabular}{c}
\psfig{file=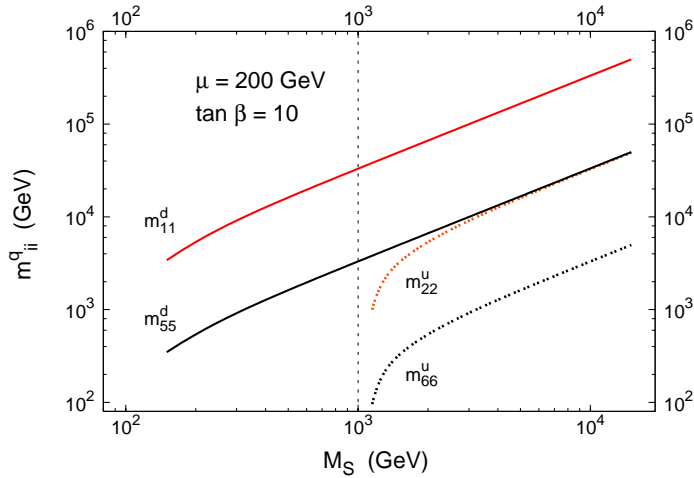,angle=270,width=100mm,clip=}
    \end{tabular}
    \caption{Minima solutions for non-degenerate VEVs as a function of the 
      SUSY scale $M_S$, with $\mu=200$ GeV, and $\tan \beta=$10.
      To the left of the vertical dotted line, the solutions for $m^q_{ii}$
      are not true minima of the potential.}
    \label{fig:mh2out.msusy.k2}
  \end{center}
\end{figure}

\begin{figure}[t]   
\begin{center}
\psfig{file=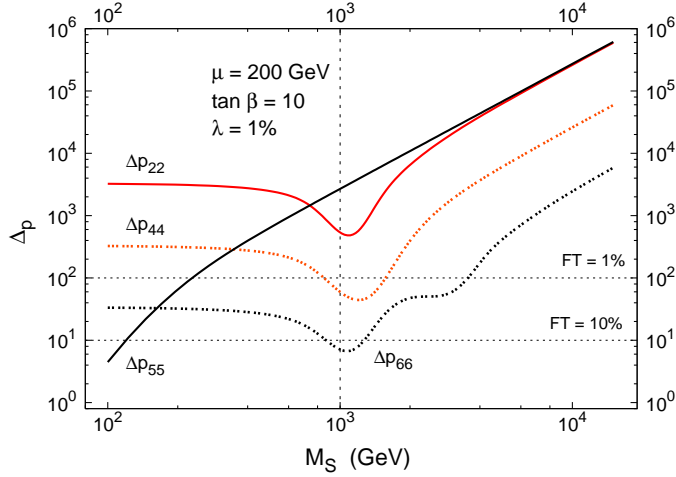,angle=270,width=100mm,clip=}
    \caption{$\Delta p$ as a function of the SUSY scale $M_S$
    for perturbations of 1\% in the diagonal soft masses, for non-degenerate 
    VEVS. We take $\mu=$200 GeV and $\tan \beta=$10.
      To the left of the vertical dotted
    line, the solutions for $m^q_{ii}$ are not true minima of the potential.
    Horizontal lines denote areas above which
    one roughly expects a FT stronger than 1\% and 10\%.}
    \label{fig:Dpout.msusy.k2}
  \end{center}
\end{figure}

\noindent
In this case, and when compared to the degenerate one,
each soft mass exhibits a very distinct behaviour at the minima of the
potential. We will thus analyse the solutions which in each case are
associated with the lightest and heaviest VEV, $(m_d)^2_{11}$,
$(m_u)^2_{22}$ and $(m_d)^2_{55}$, $(m_u)^2_{66}$.
As before, in Fig.~\ref{fig:mh2out.msusy.k2} we
display the above masses as a function of $M_S$ for $\mu=200$
GeV, and $\tan \beta=$10, similar to what was presented for degenerate
VEVs in Fig.~\ref{fig:mh2out.msusy.k1}. 
As seen from Fig.~\ref{fig:mh2out.msusy.k2}, 
the range for the solutions is not strongly
affected. Notice however that the minima values of 
$(m_u)^2_{22}$ and $(m_d)^2_{55}$ become degenerate for large values
of $M_S$. Moreover, the soft mass associated with the largest VEV
($m^u_{66}$) presents, as expected, much smaller values than the
others. This is a consequence of the suppression of $w_j/w_6$ on $B
\mu$.
Regarding the fine tuning, the situation is more involved.
In Fig.~\ref{fig:Dpout.msusy.k2} we plot $\Delta p$ for the up and down 
sectors, considering only the effect of a $\lambda^{d,u}=\lambda=
1\%$ perturbation in $(m^u_{22})^2$, $(m^u_{44})^2$, $(m^d_{55})^2$
and $(m^u_{66})^2$. We take $\mu=200$ GeV and $\tan \beta=10$.

As seen from Fig.~\ref{fig:Dpout.msusy.k2}, the alterations emerge in both
up- and down-type Higgs sectors. Regarding the down-type scalar
($\Delta p_{55}$) the fine tuning is typically larger now, and 
the effect is essentially due to the fact that it is now much more
difficult to find cancellations among the several terms entering in
the minimisation equations.
In the up-type sector, we find a very similar situation to that of 
Fig.~\ref{fig:Dpout.msusy.k1} for the second and third generations (associated
with the larger VEVs). Conversely, and since it is now associated to
a tiny VEV, the first generation up-type diagonal soft mass is
more unstable under perturbations, and its behaviour merges
with that of $(m^d_{55})^2$ for larger values of $M_S$.

It has been shown in the MSSM that with a soft SUSY
scale of a few hundred GeV the associated FT is around the
level of 10\%~\cite{Casas:2004gh}.
When compared to the MSSM, this extended model offers a 
more problematic FT scenario, as is manifest in 
Figs.~\ref{fig:Dpout.msusy.k1} and~\ref{fig:Dpout.msusy.k2}.
This can also be seen from the comparison of Eq.~(\ref{md11:min}) 
with its MSSM counterpart. In addition to the
single $(B\mu)_{12}$ and $\mu^2_{12}$ terms, one now encounters additional 
$(B\mu)_{1j}$ and $\mu_{1l}\mu_{il}$, as well as new soft breaking
masses on the right hand-side of the above equation, 
which must cancel out to achieve the
correct value of $M_Z$.

\begin{figure}[t]   
\begin{center}
\begin{tabular}{cc}\hspace*{-12mm}
\psfig{file=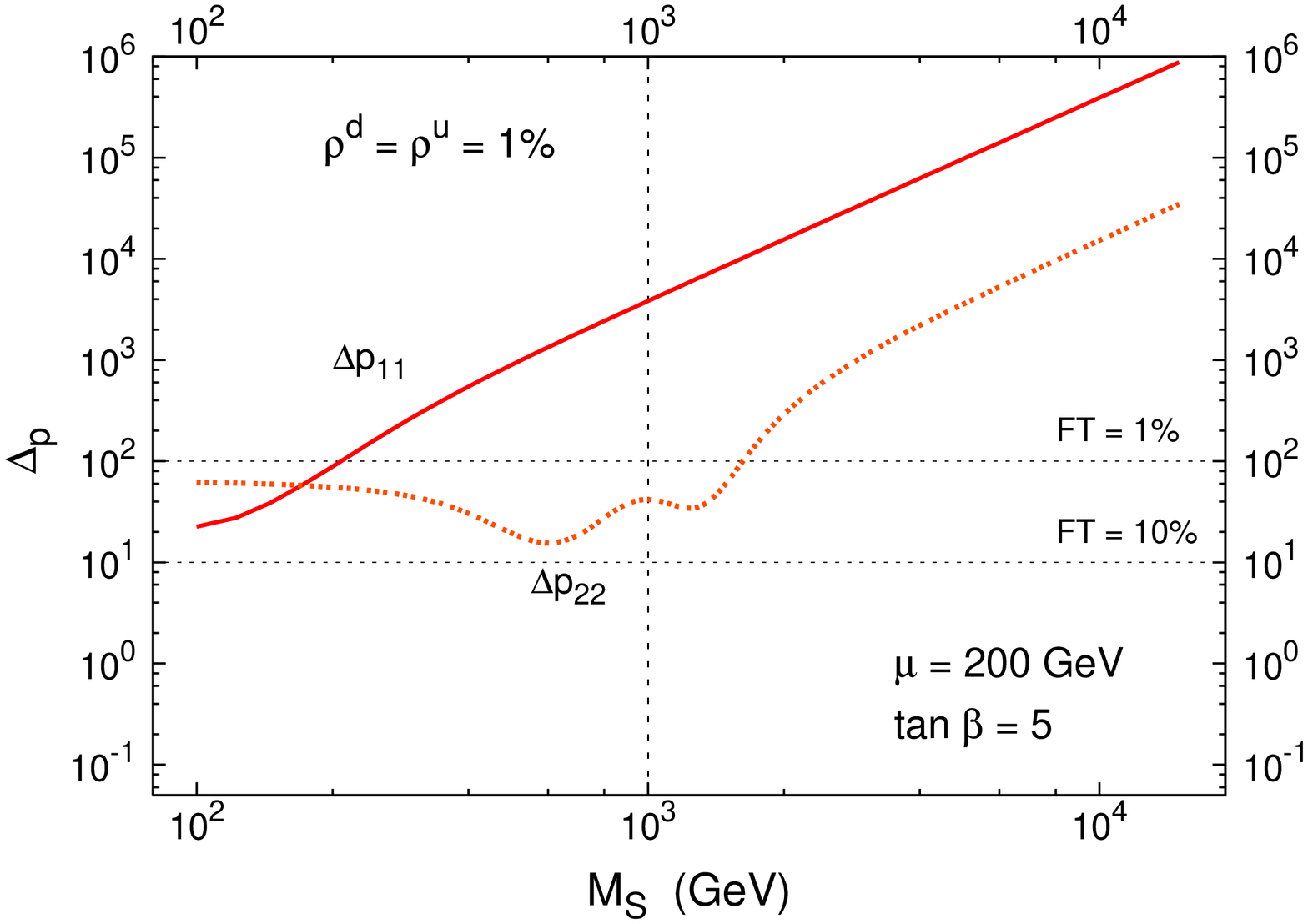,angle=270,width=90mm,clip=}
\hspace*{-5mm}&\hspace*{-7mm}
\psfig{file=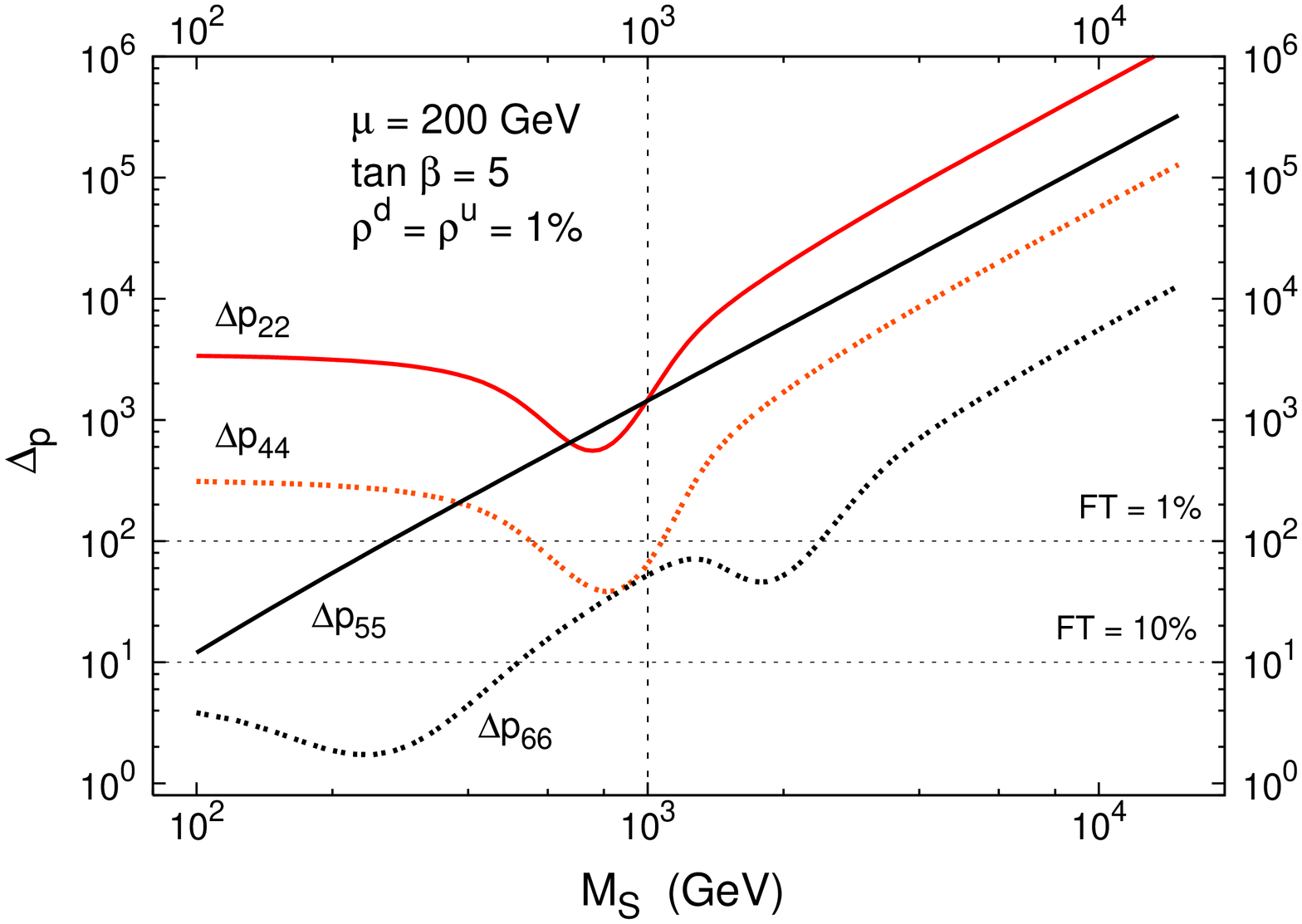,angle=270,width=90mm,clip=}
\end{tabular}
    \caption{$\Delta p$ as a function of the common SUSY scale $M_S$
    for perturbations of 1\% in $w_i$, for $\mu=$200 GeV and
    $\tan \beta=10$. On the left we present the degenerate VEV case, on the 
      right the non-degenerate case.}
    \label{fig:Dvout.msusy.k2}
  \end{center}
\end{figure}

Finally, and even though in our analysis the VEVs were
not fit by the minima equations, but taken as input
parameters instead,  
let us consider the effect of imposing a small perturbation
on the VEVs, while keeping the other parameters as defined either by
input values and/or minima conditions.
As for the previous case of perturbing the minima values of
$m^q_{ii}$, we now impose that 
\begin{equation}\label{Dw:def}
w_{i}\, \to\, 
w_{i}^\prime \,= \,(1+\rho^q)\, w_{i}\,.
\end{equation}
In Fig.~\ref{fig:Dvout.msusy.k2}, we plot the effect on $M_Z$
(parametrised by $\Delta p$) arising from taking 
$\rho^d=\rho^u=\rho=1\,\%$ for the case of degenerate (left) and
non-degenerate VEVs (right). 
Again we consider $\mu=$ 200 GeV and $\tan \beta=10$.igs/

It suffices to comment that the larger the VEVs, the more stable they are
under perturbations, and this effect is particularly manifest 
for the non-degenerate VEV case.

\section{Tree-level Higgs mass matrices}\label{higgsspectrum}
We are now in conditions to investigate the Higgs spectrum, which
will contain eleven neutral states, and ten charged physical particles. 
Let us begin with the derivation of the tree level mass matrices for
the charged scalars. 

\subsection{Charged Higgses}
In the basis defined by\footnote{The rotated basis of 
the charged fields is obtained via an identical transformation 
of Eq.~(\ref{higgs:Ptransf}).}
($\phi^-_1,\phi^{+*}_2,\phi_3^-,\phi_4^{+*},\phi_5^-,
\phi_6^{+*}$),
the scalar potential for the charged components reads,
\begin{align} 
V^{\pm} = & 
\sum_{i=1,3,5} m_i^2 \, |\phi_i^-|^2 + 
\sum_{i=2,4,6} m_i^2 \, |\phi_i^+|^2 + \frac{g^2+g'^2}{8}  \,
\sum_{i=1}^6 \, |H_i|^2 \,|H_i|^2 + 
\frac{g^2}{4}  \,\sum_{\substack{i,j=1\\i\neq j}}^6 |\dg{H_i} \, H_j|^2  
\nonumber \\
& - \frac{g^2+g'^2}{8} \sum_{\substack{i,j=1\\i+j=\text{odd}}}^6
|H_i|^2  \,|H_j|^2 - \, \frac{g^2-g'^2}{8}
\sum_{\substack{i,j=1\\i+j=\text{even}\\i\neq j}}^6
|H_i|^2 \, |H_j|^2 + (m_{13}^2  \, \phi_1^{-*} \, \phi_3^-  \nonumber \\
& + m_{15}^2 \, \phi_1^{-*} \, \phi_5^- + m_{35}^2  \,\phi_3^{-*} \, 
\phi_5^- + 
m_{24}^2  \,\phi_2^{+*}  \,\phi_4^+ + m_{26}^2  \,\phi_2^{+*}  \,\phi_6^+ + 
m_{46}^2 \, \phi_4^{+*}  \,\phi_6^+ \nonumber \\
& + \mathrm{H.c.}) + 
(b_{12} \, \phi_1^- \, \phi_2^+ + b_{14} \, \phi_1^- \, \phi_4^+ + 
b_{16} \, \phi_1^-  \,\phi_6^+ + b_{32}  \,\phi_3^-  \,\phi_2^+ + 
b_{34} \, \phi_3^-  \,\phi_4^+ \nonumber \\
& + b_{36} \, \phi_3^-  \,\phi_6^+ + b_{52} \, \phi_5^- \, \phi_2^+ + 
b_{54}  \,\phi_5^- \, \phi_4^+ + b_{56} \, \phi_5^-  \,\phi_6^+ + 
\mathrm{H.c.}) \,.
\end{align}
At the minimum of the potential, the mass matrix for the charged
states is:
\begin{equation}\label{mass:ch}
\mathcal{M}_{\pm}^2=\left( \begin{array}{cccccc}
\ssty{b_{12}\tan\beta+M_W^2\sin^2\beta} &
\ssty{b_{12}+\frac{1}{2}M_W^2\sin 2\beta} & \ssty{b_{32}\tan\beta} &
\ssty{b_{14}} & \ssty{b_{52}\tan\beta} & \ssty{b_{16}} \\
\ssty{b_{12}+\frac{1}{2}M_W^2\sin 2\beta} &
\ssty{b_{12}\cot\beta}+M_W^2\cos^2\beta & \ssty{b_{32}} &
\ssty{b_{14}\cot\beta} & \ssty{b_{52}} & \ssty{b_{16}\cot\beta} \\
\ssty{b_{32}\tan\beta} & \ssty{b_{32}} & \ssty{m_{33}^2-T_W} & \ssty{b_{34}} &
\ssty{m_{35}^2} & \ssty{b_{36}} \\
\ssty{b_{14}} & \ssty{b_{14}\cot\beta} & \ssty{b_{34}} & \ssty{m_{44}^2+T_W}  &
\ssty{b_{54}} & \ssty{m_{46}^2} \\
\ssty{b_{52}\tan\beta} & \ssty{b_{52}} & \ssty{m_{35}^2} & \ssty{b_{54}} &
\ssty{m_{55}^2-T_W} & \ssty{b_{56}} \\
\ssty{b_{16}} & \ssty{b_{16}\cot{\beta}} & \ssty{b_{36}} & \ssty{m_{46}^2} &
\ssty{b_{56}} & \ssty{m_{66}^2+T_W}
\end{array} \right)\,,
\end{equation}
where $M_W$ is the mass of the $W^\pm$ boson and
$T_W=\frac{1}{2}M_W^2\cos(2\beta)(1-\tan^2\theta_W)$. One can easily
identify the massless combinations that will give rise to the charged
Goldstone bosons ($G^{\pm}$), ``eaten away'' as the $W^\pm$ acquire
mass. The eigenstates will be henceforth denoted by $h_i^\pm$, with
$i=1$ corresponding to the unphysical massless state. 
The upper-left $4 \times 4$ sub-matrix of the one above
displayed is similar in structure 
to that derived in~\cite{Drees:1988fc}, in the framework of a
four-Higgs doublet model.

We stress here that the compact appearance of the matrix in
Eq.~(\ref{mass:ch}) is a direct consequence of working in the Higgs basis.
From direct inspection of the above matrix, one can impose that the
diagonal blocks have semi-positive eigenvalues, thus deriving a zeroth
order condition for avoiding tachyonic charged states. For example,
for the $(\phi_3^- - \phi_4^+)$ sector, this would read:
\begin{equation}
m^2_{33}\,m^2_{44} \gtrsim b^2_{34}\,,
\end{equation}
while an identical argumentation within a sector (e.g.$(\phi_3^- -
 \phi_5^-)$ in the down-type
Higgs) would in turn lead to 
\begin{equation}
m^2_{33}\,m^2_{55} \gtrsim m^4_{35}\,.
\end{equation}
Albeit very useful, the above equations are only necessary (rather
than sufficient) conditions to ensure the presence of true minima, so
they should be interpreted as a means of orientation in the parameter
space. Additionally, we notice that simple conditions as those above
are only possible to derive in the Higgs basis.

\subsection{Neutral Higgses}
In our work, we are particularly 
interested in the neutral Higgs bosons. In order to find
the neutral spectrum of the model, we decompose the complex fields as
in Eq.~(\ref{higgsbasis:decomp}), and derive the mass matrices
associated with the real and imaginary components.
CP conservation in the Higgs sector translates in the absence of terms
mixing the $R_i$ and $I_i$ components, so that scalar and pseudoscalar
states do not mix. For the scalar mass matrix we obtain
\begin{equation}
\mathcal{M}_{R}^2=
\frac{1}{2}\frac{\partial^2V}{\partial R_i\partial R_j}\Bigg|_{\text{min}} 
\nonumber
\end{equation}
\begin{displaymath}
\scriptstyle{=} \left( \begin{array}{cccccc}
\ssty{b_{12} \tan \beta + M_Z^2 \cos^2 \beta} 
& \ssty{-b_{12}-\frac{M_Z^2}{2}\sin 2\beta } 
& \ssty{b_{32}\tan\beta} & \ssty{-b_{14}} 
& \ssty{b_{52}\tan\beta} & \ssty{-b_{16}} \\
\ssty{-b_{12}-\frac{M_Z^2}{2}\sin 2\beta } 
& \ssty{b_{12} \cot \beta  + M_Z^2 \sin^2 \beta } 
& \ssty{-b_{32}} & \ssty{b_{14}\cot\beta} 
& \ssty{-b_{52}} & \ssty{b_{16}\cot\beta} \\
\ssty{b_{32}\tan\beta} & \ssty{-b_{32}} 
& \ssty{m_{33}^2+\frac{M_Z^2}{2}\cos 2\beta } 
& \ssty{-b_{34}} & \ssty{m_{35}^2} & \ssty{-b_{36}} \\
\ssty{-b_{14}} 
& \ssty{b_{14}\cot\beta} 
& \ssty{-b_{34}} 
& \ssty{m_{44}^2-\frac{M_Z^2}{2}\cos 2\beta} 
& \ssty{-b_{54}} 
& \ssty{m_{46}^2} \\
\ssty{b_{52}\tan\beta} 
& \ssty{-b_{52}} 
& \ssty{m_{35}^2} 
& \ssty{-b_{54}} 
& \ssty{m_{55}^2+\frac{M_Z^2}{2}\cos 2\beta} 
& \ssty{-b_{56}} \\
\ssty{-b_{16}} 
& \ssty{b_{16}\cot{\beta}} 
& \ssty{-b_{36}} 
& \ssty{m_{46}^2} 
& \ssty{-b_{56}} 
& \ssty{m_{66}^2-\frac{M_Z^2}{2}\cos 2\beta }
\end{array} \right)\,.
\end{displaymath}
\begin{equation}\label{RI:scalar:mx}
\end{equation}
It is straightforward to
recognise the MSSM scalar Higgs mass matrix as the $2\times 2$ upper
left block of the previous $6 \times 6$ matrix.
Likewise, the tree-level mass matrix for the pseudoscalar can be
written as 
\begin{equation}
\mathcal{M}_{I}^2=
\frac{1}{2}\frac{\partial^2V}{\partial I_i\partial I_j}\Bigg|_{\text{min}} 
\nonumber
\end{equation}
\begin{displaymath}
= \left( \begin{array}{cccccc}
\ssty{b_{12}\tan\beta} & \ssty{b_{12}} & \ssty{b_{32}\tan\beta} & 
\ssty{b_{14}} & \ssty{b_{52}\tan\beta} & \ssty{b_{16}} \\
\ssty{b_{12}} & \ssty{b_{12}\cot\beta} & \ssty{b_{32}} & 
\ssty{b_{14}\cot\beta} & \ssty{b_{52}} & \ssty{b_{16}\cot\beta} \\
\ssty{b_{32}\tan\beta} & \ssty{b_{32}} & 
\ssty{m_{33}^2+\frac{M_Z^2}{2}\cos 2\beta} & \ssty{b_{34}} & 
\ssty{m_{35}^2} & \ssty{b_{36}} \\
\ssty{b_{14}} & \ssty{b_{14}\cot\beta} & \ssty{b_{34}} & 
\ssty{m_{44}^2-\frac{M_Z^2}{2}\cos 2\beta} & \ssty{b_{54}} & \ssty{m_{46}^2} \\
\ssty{b_{52}\tan\beta} & \ssty{b_{52}} & \ssty{m_{35}^2} & 
\ssty{b_{54}} & \ssty{m_{55}^2+\frac{M_Z^2}{2}\cos2\beta} & \ssty{b_{56}} \\
\ssty{b_{16}} & \ssty{b_{16}\cot{\beta}} & \ssty{b_{36}} & 
\ssty{m_{46}^2} & \ssty{b_{56}} & \ssty{m_{66}^2-\frac{M_Z^2}{2}\cos 2\beta}
\end{array} \right)\,.
\end{displaymath}
\begin{equation}\label{RI:pseudoscalar:mx}
\end{equation}
In the $2\times2$ submatrix defined by the $i,j=1,2$ entries, it
is easy to identify the combination associated with the massless
Goldstone boson. As in the MSSM, the latter degree of freedom is
``eaten'' by the $Z^0$ boson, as it acquires a mass. After rotating
away the massless state,
\begin{equation}
G^0=\frac{1}{\sqrt{2}}(I_2\sin\beta-I_1\cos\beta)\,,
\end{equation}
by means of the unitary transformation
\begin{equation}\label{rotate:pseudo}
\left( \begin{array}{c}
I_1' \\
I_2'
\end{array} \right)=
\left( \begin{array}{cc}
\cos{\beta} & -\sin{\beta} \\
\sin{\beta} & \cos{\beta}
\end{array} \right)\left( \begin{array}{c}
I_1 \\
I_2
\end{array} \right) \,,
\end{equation}
the five remaining eigenstates are in general massive, and the new
mass matrix reads
\begin{equation}
\mathcal{M}_{I}^{'2}=\left( \begin{array}{ccccc}
\sty{\frac{2b_{12}}{\sin 2\beta}} & \sty{b_{32}\sec{\beta}} & 
\sty{b_{14}\csc\beta} & \sty{b_{52}\sec\beta} & \sty{b_{16}\csc\beta} \\
\sty{b_{32}\sec\beta} & \sty{m_{33}^2+\frac{m_Z^2}{2}\cos 2\beta } & 
\sty{b_{34}} & \sty{m_{35}^2} & \sty{b_{36}} \\
\sty{b_{14}\csc\beta} & \sty{b_{34}} & 
\sty{m_{44}^2-\frac{m_Z^2}{2}\cos 2\beta} & \sty{b_{54}} & \sty{m_{46}^2} \\
\sty{b_{52}\sec\beta} & \sty{m_{35}^2} & \sty{b_{54}} & 
\sty{m_{55}^2+\frac{m_Z^2}{2}\cos 2\beta } & \ssty{b_{56}} \\
\sty{b_{16}\csc\beta} & \sty{b_{36}} & \sty{m_{46}^2} & 
\sty{b_{56}} & \sty{m_{66}^2-\frac{m_Z^2}{2}\cos 2\beta }
\end{array} \right)\,.\label{RI:pseudoscalar:mx2}
\end{equation}
When discussing the issue of minimising the scalar potential, we
pointed out that each of the presented basis (Higgs and superpotential basis)
had its own advantages/drawbacks. Here we are facing a very strong
point in favour of the Higgs basis, namely the possibility of rotating
away the pseudoscalar massless state via a transformation that only
involves the first two states. Should we have been working in the
superpotential basis, 
the transformation of Eq.~(\ref{rotate:pseudo}) would be far
more complex. Even though one could still define $\tan \beta$ as 
$\tan \beta = \sqrt{w_2^2+w_4^2+w_6^2} \,/  \,\sqrt{w_1^2+w_3^2+w_5^2}$, the
rotation parametrised by $\beta$ would be extended to a $6 \times 6$
matrix, which would act upon the combination of $P^\dagger \, (I_1, I_2,
\dots I_6)^T$.

It is worth recalling that the above mass matrices (scalar and
pseudoscalar) are not those
associated with the original (interaction) eigenstates. The relation
between the superpotential basis and the Higgs basis is trivially obtained 
from Eq.~(\ref{higgs:Ptransf}),
\begin{equation}
R_i = P_{ij} \sigma_j\,, \quad \quad
I_i = P_{ij} \varphi_j\,, \quad \quad
i,j = 1 \ldots 6\,.
\end{equation}
In the Higgs basis, the mass matrices can be easily diagonalised by
\begin{align}\label{higgsmass:diag}
S_R \mathcal{M}^2_R S_R^\dagger &= \Delta_R^2 = 
\operatorname{diag}({m^s_i}^2)\,, 
\nonumber\\
S_I \mathcal{M}^{2}_I S_I^\dagger &= \Delta_I^2=
\operatorname{diag}({m^p_i}^2)\,,
\end{align}
where $\Delta_{R,I}^2$ are the diagonal scalar and pseudoscalar
squared mass eigenvalues (notice that the $i=1$ term for the
pseudoscalars corresponds to the unphysical massless would-be
Goldstone boson). From the above it is straightforward to observe that 
the matrices that diagonalise the mass matrices in the original basis
can be related to the latter as 
\begin{equation}\label{Ssigma:SP}
S_{\sigma,\varphi} \,=\,S_{R,I} \,P\,. 
\end{equation}

\subsection{Tree-level Higgs spectrum - a brief discussion}

Although a thorough analysis of the Higgs parameter space lies beyond
the scope of this paper, it is important to comment on a few issues.
The first regards radiative corrections, which play a key role in the
MSSM and its extensions, especially in relation with the mass of the
lightest scalar Higgs. 
Without radiative corrections, and as occurs in the MSSM, 
the mass of the lightest scalar is bounded to be $m_{h^0_1} \lesssim
M_Z$~\cite{Flores:1982pr}. 
In the framework of supersymmetric models with extended Higgs sectors,
higher order contributions to the lightest Higgs mass have been
already analysed~\cite{Sakamura:1999ky}, and the upper bound on $m_{h^0_1}$
does not differ significantly from the one derived in the MSSM.
Since in this work our aim is not so much to study in depth the Higgs
spectra, but rather investigate to which extent FCNCs push the lower
bounds on the heavy Higgs masses, we simplify the analysis, and use
the bare masses instead. 

From the analysis of the tree-level mass matrices derived in this Section,
Eqs.~(\ref{mass:ch},\ref{RI:scalar:mx},\ref{RI:pseudoscalar:mx},\ref{RI:pseudoscalar:mx2}),
and again working in the Higgs basis, a few interesting patterns can
be extracted for the behaviour of the several mass eigenstates.
Beginning with the lighter physical states ($i=2$ for both charged and
pseudoscalar states), the bare masses are very similar to what one has in
the MSSM:
\begin{align}
& m^s_1 \,\lesssim \,\left| \,\cos 2 \beta\, \right| \,M_Z\,, \nonumber \\
&(m^s_2)^2 \,\approx \,(m^p_2)^2\, +\, M_Z^2\,,\nonumber \\
&(m^p_2)^2 \,\simeq \,2 \,\frac{b_{12}}{\sin 2 \beta}\,,\nonumber \\
&(m^\pm_2)^2 \,\approx \,(m^p_2)^2\, +\, M_W^2\,.
\end{align}
In the above, the approximations encode the fact that we have
neglected the mixing between the several states (in each scalar, pseudoscalar
and charged sector). Taking the mixing into account would lead to
cumbersome expressions, involving all the states, which could only be
numerically evaluated. 
For the remaining (heavier) states the physical masses are
dominated to a large extent by the diagonal entries in the
corresponding mass matrices, i.e., $m^2_{33-66}$. Although the
general case is that the presence of the off-diagonal terms
($m^2_{ij}$ and $b_{ij}$) does not substantially modify the mass
spectrum, the situation can be distinct if either one of the latter
quantities becomes close to the values of the diagonal entries, namely
\begin{equation}
m^2_{ij}\, \approx \,0.9 \,\,m^2_{ii}\,, \quad \quad \text{or} \quad \quad 
b_{ij}\, \approx \,0.35 \,\,m^2_{ii}\,. 
\end{equation}
If we are in the presence of the above regimes, then more mixing
between the states is produced, and one is ultimately led to the
appearance of tachyons. 
The presence of tachyons is also triggered by increasingly larger
values of $\tan \beta$.

Even though we will not address experimental issues, we will adopt the
following na\"{\i}ve bounds (mimicking the MSSM) 
for the bare masses of the lightest
states~\cite{pdg2004}:
\begin{align}\label{higgs:bound}
m^s_1 \,\gtrsim \,75 \,\,\text{GeV}\,, \quad \quad 
m^p_2 \,\gtrsim \,91 \,\,\text{GeV}\,,\quad \quad 
m^\pm_2 \,\gtrsim \,80 \,\,\text{GeV}\,,
\end{align}
where the first bound translates in taking $\tan \beta > 3$.

\section{Yukawa interaction Lagrangian}\label{yukint}
In agreement with the superpotential introduced in Section~\ref{higgsphenom}, 
the Lagrangian for the interaction of Higgs with up- and down-quarks 
can be written as 
\begin{align}
\mathcal{L}^0_{\text{Yukawa}}=
-\operatornamewithlimits{\sum}_{i=1,3,5} 
\, h_i^0 \, \bar d_{R}^\prime \, Y^d_i \,   d_{L}^\prime 
-\operatornamewithlimits{\sum}_{i=2,4,6} 
\, h_i^0 \, \bar u_{R}^\prime \, Y^u_i \, u_{L}^\prime  +\text{H.c.}\,,
\end{align}
where $q_{L,R}$ ($Y^{u,d}_i$) are vectors (matrices) in flavour space,
and the quarks appearing above are interaction (rather than mass)
eigenstates, a prime being used to emphasise the latter.
This Lagrangian gives rise to the quark mass matrices 
and to scalar and pseudoscalar Higgs-quark-quark
interactions.
The mass terms for the quarks read
\begin{align}\label{massL}
\mathcal{L}_{\text{mass}}=& 
-
\bar d_R^\prime \, M^d \, d_L^\prime 
-
\bar u_R^\prime \, M^u \, u_L^\prime  +
\text{H.c.}\,, \\
& \quad M^d=\operatornamewithlimits{\sum}_{i=1,3,5} w_i \, Y^d_i
\,, \quad \quad
M^u=\operatornamewithlimits{\sum}_{i=2,4,6} w_i \, Y^u_i
\,.
\end{align}
The resulting up- and down-type quark mass matrices can be
diagonalised as 
\begin{equation}\label{quarkmass:diag}
V_R^q \, M^q  \,{V_L^q}^\dagger \,=\, \operatorname{diag}(m^q_i)\,, 
\quad \quad q=u,d\,,
\end{equation}
while the mass and interaction (primed) eigenstates are related by the
following transformations
\begin{equation}
V_L^q \, q_L^\prime = q_L \, \quad \quad
V_R^q \, q_R^\prime= q_R \quad \quad q=u,d\,.
\end{equation}
The Cabibbo-Kobayashi-Maskawa (CKM) matrix is defined as 
\begin{equation}\label{vckm:def}
V_{\text{CKM}}=V_L^u  \,{V_L^d}^\dagger\,.
\end{equation}
From the above equations it is manifest that in the quark sector, mass
matrices and Yukawa couplings are in general misaligned, the only
exception occurring for Yukawa couplings which are proportional.
This misalignment translates into the impossibility of
diagonalising both matrices simultaneously, and is the source of the
existence of tree-level FCNCs, as we will briefly discuss.

Let us turn our attention to the Higgs-quark-quark interaction Lagrangian. 
In the unrotated basis, the latter reads
\begin{align}
\mathcal{L}^0_{\text{Yukawa}}=& 
-\frac{1}{\sqrt{2}}
\operatornamewithlimits{\sum}_{i=1,3,5} \left[\,
\sigma_i \,\bar d_{R}^\prime \, Y^d_i\,  d_{L}^\prime +
i \varphi_i \,\bar d_{R}^\prime \, Y^d_i \,d_{L}^\prime + \text{H.c.}
\right]\nonumber\\
&
-\frac{1}{\sqrt{2}}
\operatornamewithlimits{\sum}_{i=2,4,6} \left[\,
\sigma_i \, \bar u_{R}^\prime \,Y^u_i \, u_{L}^\prime +
i \varphi_i\, \bar u_{R}^\prime \,Y^u_i \, u_{L}^\prime
+\text{H.c.}\right]\,, 
\end{align}
while moving to the quark and Higgs mass eigenstate basis the couplings are 
\begin{align}
\mathcal{L}_{\text{Yukawa}}=&  
-\frac{1}{\sqrt{2}}
\operatornamewithlimits{\sum}_{i=1,3,5} \left[\,
\left(\mathcal{V}_d\right)^{ij}_{ab} \,
h_j^s \, \bar d_{R}^a \,d_{L}^b + 
i \left(\mathcal{W}_d\right)^{ij}_{ab} \,h_j^p \,\bar d_{R}^a \,d_{L}^b +
\text{H.c.} \right]\nonumber\\
&
-\frac{1}{\sqrt{2}}
\operatornamewithlimits{\sum}_{i=2,4,6} \left[
\left(\mathcal{V}_u\right)^{ij}_{ab} \,
h_j^s \, \bar u_{R}^a \,u_{L}^b + 
i \left(\mathcal{W}_u\right)^{ij}_{ab} \,h_j^p \,\bar u_{R}^a \,u_{L}^b +
\text{H.c.} \right]\,.
\end{align}
In the above, $a,b$ denote quark flavours, while
$i,j=1, \ldots, 6$ are Higgs indices, with $s$ ($p$)
denoting scalar (pseudoscalar) mass eigenstates. The latter are related
to the original states as $h^s=S_\sigma \sigma$, $h^p=S_\varphi
\varphi$, as from Eqs.~(\ref{higgsmass:diag},\ref{Ssigma:SP}). 
The scalar (pseudoscalar) couplings 
$\mathcal{V}$ ($\mathcal{W}$) are defined as
\begin{align}\label{WV}
\left(\mathcal{V}_q\right)^{ij}_{ab} =&
(S_\sigma^\dagger)_{ij} \, \,
(V_R^q\, Y^{q}_i \,V^{q\dagger}_L)_{ab}\,, \nonumber \\
\left(\mathcal{W}_q\right)^{ij}_{ab} =&
(S_\varphi^\dagger)_{ij}\, \, 
(V_R^q \, Y^{q}_i \,V^{q\dagger}_L)_{ab}\,,
\end{align}
with $i=1,3,5 \ (2,4,6)$ for $q=d \,(u)$ and $j=1, \ldots, 6$.
The several rotation matrices
appearing in the previous equation were defined in 
Eqs.~(\ref{higgsmass:diag},\ref{Ssigma:SP},\ref{quarkmass:diag}).
As a final remark, it is important to stress that the matrices 
$V_{L,R}$ which diagonalise the quark mass matrices do not, in
general, diagonalise the corresponding Yukawa couplings. Hence, both
scalar and pseudoscalar Higgs-quark-quark interactions may exhibit a
strong non-diagonality in flavour space.

Finally, let us point out that in the Higgs basis 
$R_1$ and $R_2$ do not have flavour violating interactions. This is
straightforward if one recalls that the Higgs basis mimics an
MSSM-extended model. In the basis of the Higgs physical states,
all six (five) scalars (pseudoscalars) mix, and all play a role in
mediating the FCNC processes.

\section{FCNCs at the tree-level}\label{fcnc}
In this section, we compute the tree-level observables (such as
neutral meson mass differences and CP violation in neutral meson
mixing) induced by the exchange of neutral Higgs. These effects are
absent in the SM, and play a determinant role in constraining the free
parameters of the model.
We discuss these effects for the case of the neutral kaons,
as well as for the $B_d$, $B_s$ and $D^0- \bar D^0$ systems.

\subsection{$K$-meson oscillations and contributions to $\Delta
  m_K$}\label{kaons}
We begin with the computation of the contributions of the several Higgs
fields to the mass difference of the long- and short-lived neutral
kaon states. In terms of effective Hamiltonians, the neutral kaon mass
difference is defined as 
\begin{equation}
\Delta m_K = m_{K_L} - m_{K_S} \simeq 2 \left| \mathcal{M}^K_{12}\right|=
2 \left| \langle \overline{K}^0 \left|
\mathcal{H}^{\Delta S=2}_{\text{eff}}\right| K^0 \rangle\right|\,,
\end{equation} 
where $\mathcal{H}^{\Delta S=2}_{\text{eff}}$ is the effective
Hamiltonian governing $\Delta S=2$ transitions. The Hamiltonian can be
decomposed as 
\begin{equation}
\mathcal{H}^{\Delta S=2}_{\text{eff}} = \mathcal{H}_{\text{tree}}+
\mathcal{H}_{\text{loop}}\,.
\end{equation}
In the above we have separated the contributions arising from
tree-level diagrams from those associated with box and higher-loop
diagrams. We will focus on the tree-level contributions to 
$\Delta m_K$ induced by the exchange of scalar and pseudoscalar
Higgs. 
\begin{figure}[t]
  \begin{center}
    \begin{tabular}{cc}
      \epsfig{file=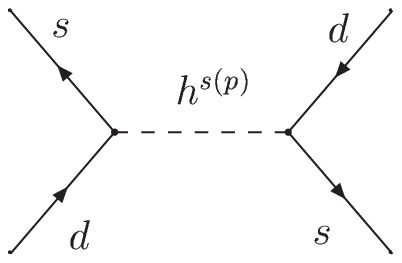,clip=}
      \hspace*{12mm}&\hspace*{12mm}
      \epsfig{file=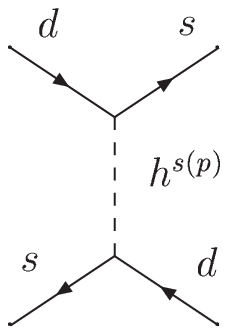,clip=}
      \\ & \\
      (a)\hspace*{12mm}&\hspace*{12mm} (b)
    \end{tabular}
    \caption{Feynman diagrams contributing to $\Delta m_K$ at
      tree-level.}
    \label{fig:effH}
  \end{center}
\end{figure}
From the interaction Lagrangians previously derived, it is now
straightforward to compute the effective Hamiltonian for the
diagrams in Fig.~\ref{fig:effH}. One thus has
$\mathcal{H}_{\text{eff}}=\mathcal{H}^\sigma_{\text{eff}}+ 
\mathcal{H}^\varphi_{\text{eff}}$, with
\begin{align}
\mathcal{H}^\sigma_{\text{eff}}=&
\operatornamewithlimits{\sum}_{\begin{smallmatrix}
{j=1-6}
\end{smallmatrix}} \frac{1}{16 (m^s_j)^2}\left\{
\bar s \, \operatornamewithlimits{\sum}_{i=1,3,5}\left[ 
({\mathcal{V}_d^\dagger}+\mathcal{V}_d)^{ij}_{21}+
({\mathcal{V}_d^\dagger}-\mathcal{V}_d)_{21}^{ij}\gamma_5
\right]d
\right\}^2 +\text{H.c.}\,, \label{KK:effH:sigma}\\
&\nonumber\\
\mathcal{H}^\varphi_{\text{eff}}=&
-\operatornamewithlimits{\sum}_{\begin{smallmatrix}
{j=2-6}
\end{smallmatrix}} \frac{1}{16 (m^p_j)^2}\left\{
\bar s \,\operatornamewithlimits{\sum}_{i=1,3,5} \left[ 
({\mathcal{W}_d}-\mathcal{W}_d^\dagger)^{ij}_{21}-
({\mathcal{W}_d}+\mathcal{W}_d^\dagger)_{21}^{ij}\gamma_5
\right]d
\right\}^2 +\text{H.c.}\,,\label{KK:effH:varphi}
\end{align}
where the scalar and pseudoscalar Higgs masses, $m^{s,p}$, have been 
defined in Eq.~(\ref{higgsmass:diag}).
Therefore, the contribution to the kaon mass difference associated
with the exchange of a scalar Higgs boson ($h^s$) is
given by 
\begin{align}\label{MK12:sigma}
\left. \mathcal{M}^K_{12}\right|^\sigma =&
\frac{1}{8} \operatornamewithlimits{\sum}_{\begin{smallmatrix}
{j=1-6}
\end{smallmatrix}} \frac{1}{(m^s_j)^2}
\left\{ \left[\operatornamewithlimits{\sum}_{i=1,3,5} \left(
{\mathcal{V}_d}^{ij*}_{12}+{\mathcal{V}_d}^{ij}_{21} \right)
\right]^2 \langle \overline K^0\left| (\bar s d) (\bar s d)
\right| K^0 \rangle \right. \nonumber\\
&+
\left.
\left[\operatornamewithlimits{\sum}_{i=1,3,5} \left(
{\mathcal{V}_d}^{ij*}_{12}-{\mathcal{V}_d}^{ij}_{21} \right)
\right]^2 \langle \overline K^0\left| (\bar s \gamma_5 d) (\bar s
\gamma_5 d)
\right| K^0 \rangle
\right\}\,, 
\end{align}
while the exchange of a pseudoscalar state ($h^p$) reads
\begin{align}\label{MK12:phi}
\left. \mathcal{M}^K_{12}\right|^\varphi =&
\frac{1}{8} \operatornamewithlimits{\sum}_{\begin{smallmatrix}
{j=2-6}
\end{smallmatrix}} \frac{1}{(m^p_j)^2}
\left\{ \left[\operatornamewithlimits{\sum}_{i=1,3,5} \left(
{\mathcal{W}_d}^{ij*}_{21}-{\mathcal{W}_d}^{ij}_{12} \right)
\right]^2 \langle \overline K^0\left| (\bar s d) (\bar s d)
\right| K^0 \rangle \right. \nonumber\\
&+
\left.
\left[\operatornamewithlimits{\sum}_{i=1,3,5} \left(
{\mathcal{W}_d}^{ij*}_{21}+{\mathcal{W}_d}^{ij}_{12} \right)
\right]^2 \langle \overline K^0\left| (\bar s \gamma_5 d) (\bar s
\gamma_5 d)
\right| K^0 \rangle
\right\}\,. 
\end{align}
The theoretical prediction for the value of $\Delta m_K$ thus obtained
should be compared with the experimental value of $(\Delta
m_K)_{\text{exp}}=3.49 \times 10^{-12}$ MeV. We stress here that in a
SM/MSSM-like scenario, the ${\mathcal{W}_d}$, ${\mathcal{V}_d}$ 
matrices entering in Eqs.~(\ref{MK12:sigma}, \ref{MK12:phi}) would
be diagonal in flavour space, and thus no tree-level FCNC would
occur. This is clear from Eqs.~(\ref{WV}), since the flavour content
of $\mathcal{W}$ and $\mathcal{V}$ is proportional to 
$(V_R^d\, Y^{d}_i \,V^{d\dagger}_L)_{ab}$. In the SM and MSSM, the
matrices that diagonalise the quark mass matrices $V_{L,R}$ also
diagonalise the Yukawa couplings, so $(V_R^d\, Y^{d}_i
\,V^{d\dagger}_L)_{ab} \propto \delta_{ab}$.
In multi-Higgs doublet models, where the underlying theory implies
that the Yukawas are proportional among themselves
(i.e. $Y^q_i \propto Y^q_j$), we also encounter a situation where
no tree-level FCNCs emerge.

Given the purpose of the analysis, we have adopted a simple approach 
regarding the computation of the meson matrix elements, using the
vacuum insertion approximation with non-renormalised operators.
Following Ref.~\cite{hme}, we have
\begin{align}
\langle \overline K^0\left| (\bar s d) (\bar s d)
\right| K^0 \rangle &= \left[ 
\frac{1}{12} - \frac{1}{12} \left(\frac{m_K}{m_s+m_d}
\right)^2 \right]\,m_K\,f_K^2\,,
\nonumber \\
\langle \overline K^0\left| (\bar s \gamma_5 d) (\bar s
\gamma_5 d)
\right| K^0 \rangle &=  \left[ 
-\frac{1}{12} + \frac{11}{12} \left(\frac{m_K}{m_s+m_d}
\right)^2 \right]\,m_K\,f_K^2\,,
\end{align}
where $m_K$ is the kaon mass, and $f_K$ the kaon decay constant. 
In Table~\ref{mesondata}, we present several relevant input parameters
for the computation of meson observables.
\begin{table}
\hspace*{10mm}
\begin{tabular}{||c|c c||c|c c||}
\hline \hline
$m_K$ & $497.6$ MeV& \cite{pdg2004} 
&$m_{D^0}$ & $1.864$ GeV& \cite{pdg2004}\\
\hline
$\Delta m_K$ & $3.49 \times 10^{-12}$ MeV& \cite{pdg2004} 
&$\Delta m_{D^0}$ & $< 46.07 \times 10^{-12}$ MeV & \cite{pdg2004} \\
\hline
$f_K$ & $159$ MeV& \cite{hme} 
&$f_D$ & $224$ MeV& \cite{lattice}\\
\hline
$m_{B_d}$ & $5.279$ GeV& \cite{pdg2004} 
&$m_{B_s}$ & $5.369$ GeV& \cite{pdg2004}\\
\hline
$\Delta m_{B_d}$ & $3.304 \times 10^{-13}$ GeV& \cite{pdg2004} 
&$\Delta m_{B_s}$ & $>94.8 \times 10^{-13}$ GeV& \cite{pdg2004}\\
\hline
$f_{B_d}$ & $215$ MeV& \cite{Aoki:2003xb} 
&$f_{B_s}$ & $245$ MeV& \cite{Aoki:2003xb}\\
\hline \hline
\end{tabular}
\caption{Numerical values used throughout the computation with the
  corresponding references. 
} \label{mesondata}
\end{table}

Before concluding the analysis of the neutral kaon sector, let us
mention that in a scenario where one has flavour violating 
neutral Higgs couplings, and given 
the most generic possibility of having a complex CKM matrix, it is
natural to expect the occurrence of indirect CP violation at the
tree-level. Although we will not pursue this issue in the following numerical
analysis, let us just stress that the contributions to 
$\varepsilon_K$ (which parametrises indirect CP violation in the kaon
sector) are given by
\begin{equation}
\varepsilon_K = -\frac{e^{i \pi/4}}{\sqrt{2}} \, 
\frac{\operatorname{Im} \left[\mathcal{M}^K_{12} 
\lambda_u^2\right]}{|\lambda_u|^2 \, \Delta m_K}\,,
\end{equation}
where $\lambda_u$ is defined from CKM elements as $\lambda_u= V^*_{us}
V_{ud}$, with $\mathcal{M}^K_{12}$ as computed in 
Eqs.~(\ref{MK12:sigma},\ref{MK12:phi}) (under the assumption of complex
Yukawa couplings). This new tree-level contribution would have to be
compatible with the SM loop contribution and the experimental bound 
$\varepsilon_K = (2.284 \pm 0.014) \times 10^{-3}$~\cite{pdg2004}.  

\subsection{Other neutral meson systems: $B_d$, $B_s$ and
  $D^0$}\label{othermesons} 
Computing the mass differences of neutral $B$ and $D$ mesons
introduces no new elements into the analysis. The approach is entirely
identical to that of the kaon system, the only difference lying in
replacing the $K^0$ $(\bar s d)$ constituent quarks of
Fig.~\ref{fig:effH} by $(\bar u c)$,  $(\bar b d)$ and $(\bar b s)$,
for $D^0$, $B_d$ and $B_s$, respectively. In each case the effective
Hamiltonian for scalar and pseudoscalar Higgs exchange reads as in
Eqs.~(\ref{KK:effH:sigma}, \ref{KK:effH:varphi}), providing that the
following replacements are done\footnote{In all cases, we are only
computing an estimate value, not taking into account neither theoretical
uncertainties (as those associated with the computation of
the matrix elements), nor experimental errors.}: 

\noindent $B_d$: $\mathcal{V}$ and $\mathcal{W}$ indices $(21) \to
(31)$.

\noindent $B_s$: $\mathcal{V}$ and $\mathcal{W}$ indices $(21) \to
(32)$.

\noindent $D$: $\mathcal{V}$ and $\mathcal{W}$ computed for the
up-sector (see Eq.~(\ref{WV})); $\mathcal{V}$ and $\mathcal{W}$
indices $(21) \to (12)$; sum over interaction eigenstates $i=1,3,5 \to
i=2,4,6$. 

In each case, the hadronic matrix elements should be also recomputed, and
the predictions for each of the above processes should
be confronted with the experimental data summarised in Table~\ref{mesondata}.
Before concluding this section, let us briefly comment on the several
observables mentioned here. 
First of all, it is widely recognised that, in models with tree-level
FCNC, the most stringent bounds are usually associated with $\Delta
m_K$.
Regarding the $B_d$ sector, it has also been argued that in the
absence of a predictive theory for the Yukawa couplings, the bounds
associated with $\Delta m_{B_d}$ should be considered as a more
reliable constraint, since they do not involve the mixing between the
first two generations~\cite{Sher:1991km}. Although we include it in
our analysis, the $B_s$ mass difference is not expected to add any new
information. In the SM, this mixing is already maximal, and the
addition of a new contribution would have little effect, the only
exception occurring if new contributions matched exactly those of
the SM, but had opposite sign, in which case a cancellation could take
place. Still, this is a very fine-tuned scenario, and hardly
significant, given the uncertainties associated with the computation.
Finally, we turn our attention to the $D^0$ mass difference. As
pointed out in~\cite{McWilliams:1980kj,Cheng:1987rs} 
and~\cite{D:burdman:datta}, models
allowing for FCNC at the tree-level may present the possibility of very
large contributions to $\Delta m_D$, thus emerging as an excellent probe
of new physics effects. The latter 
contributions are often harder to control than those
associated with, for example, $\Delta m_K$. In fact, and as 
discussed in~\cite{Cheng:1987rs}, 
the contribution of tree-level FCNCs to 
$\Delta m_D$ could even exceed 
by a factor 20 those to $\Delta
m_K$. On the other hand, mixing in the $D^0$ sector is very sensitive
to the hadronic model used to estimate the transition amplitudes, and
there is still a very large uncertainty in deriving its decay
constants, etc. Therefore, the constraints on a given model arising
from $\Delta m_D$ should not be over-emphasised.

Another interesting issue is that of rare decays. It has been argued
that, again when no theory for the full Yukawas is available,
some rare decays become very sensitive to flavour changing 
contributions induced by
Higgs exchange at the tree-level. In Ref.~\cite{Sher:1991km}, the
authors have identified that the most promising decay modes involve 
the leptonic sector, together with the $B$-mesons, and are
$\mu \to e \gamma$, $B_d \to K \mu \tau$ and $B_s \to \mu \tau$. 
We will not pursue this issue in the present work, reserving some of
the above modes (namely $\mu \to e \gamma$) for a forthcoming 
analysis~\cite{EMT:lepton}.

\section{Flavour violation in six Higgs doublet models: results and
discussion}\label{flavour:res}
 
\subsection{Yukawa couplings: the simple Fritzsch scheme}
We have carried out in the previous Section a general analysis of FCNC
contributions. This can be applied to any particular model with three
Higgs families, provided that the Yukawa couplings are known.
Lead by simplicity, and following the analysis of~\cite{Cheng:1987rs}, 
we take as an illustrative example
the so-called ``simple Fritzsch 
scheme''~\cite{Fritzsch:1977vd,Fritzsch:1999ee}. 
Taking this ansatz for each
of the Yukawa couplings presents two main advantages: it leads to mass
matrices with a fairly hierarchical structure, and allows to fit
experimental data on quark masses and mixings for a reasonable number
of free parameters. 

The ``simple Fritzsch scheme'' essentially consists in having all
the weak quark eigenstates adopt identically structured couplings,
which display suppressed flavour-changing elements for the first two
generations in each family. More explicitly one has
\begin{align}\label{fritzsch:def}
Y_i\,=\,\frac{1}{w_i} \, \left(
\begin{array}{ccc}
0 & A_i & 0\\
A_i & 0 & B_i \\
0 & B_i & C_i
\end{array}
\right) \,,
\end{align}
where $i=1,...,6$, and the entries are given by\footnote{For simplicity,
and since our main concern is to illustrate the contributions to meson
mass difference, we will assume that the Yukawa couplings are real,
so that we will not examine the contributions to CP violation observables.}
\begin{equation}\label{fritzsch:abc}
A_i = a_i  \, \sqrt{m_1^q  \,m_2^q} \,, \quad \quad
B_i = b_i  \, \sqrt{m_2^q  \,m_3^q} \,, \quad \quad
C_i = c_i  \,m_3^q\,,
\end{equation}
with $a_i$, $b_i$ and $c_i$ coefficients of order one, and $m^q_j$ the
quark masses of the $j^\text{th}$ generation and $q=d \,(u)$ for
$i=1,3,5 \,(2,4,6)$.
One further advantage of this choice, which will become more evident in the
subsequent discussion, is that under the above ansatz, the quark
masses and mixings are independent of the chosen VEV regime (although
the Yukawas are not). This means that for a given successful choice of
the parameters $a_i$, $b_i$ and $c_i$, one can study a number of
distinct VEV schemes, corresponding to different $\tan \beta$ scenarios.

Let us now present a specific numerical example. Taking the 
values for the input quark masses, $m_i^q$ entering in the ansatz of
Eq.~(\ref{fritzsch:abc}) as $m_u=3$ MeV, $m_d=6.5$ MeV, $m_c=1.25$ GeV,
$m_s=0.1$ GeV, $m_t=178$ GeV, $m_b=$4.5 GeV, and using the following
choice of coefficients
\begin{align}
&
a\,=\, \{0.45, 0.30, 0.35, 0.45, 0.20, 0.50\}\,, \nonumber \\
&
b\,=\, \{0.25, 0.40, 0.45, 0.30, 0.20, 0.35\}\,, \nonumber \\
&
c\,=\, \{0.30, 0.45, 0.30, 0.30, 0.40, 0.25\}\,, 
\end{align}
one obtains the following set of output quark masses 
\begin{align}
& m_u= 0.0042\, \,\text{GeV}\,, \quad m_c = 1.37\, \,\text{GeV}\,,
  \,\quad m_t=179.4\, \,\text{GeV} 
\nonumber \\
& m_d= 0.0073\, \,\text{GeV}\,, \quad m_s = 0.087\, \,\text{GeV}\,,
  \quad m_b=4.5 \, \,\text{GeV} 
\end{align}
and the associated mixing matrix
\begin{equation}
\left| V_{\text{CKM}} \right|\,=\, \left(
\begin{array}{ccc}
0.9742 & 0.2255 & 0.0031 \\
0.2251 & 0.9733 & 0.0438 \\
0.0129 & 0.0420 & 0.9990
\end{array}
\right)\,,
\end{equation}
which is in good agreement with experimental data. Having the relevant
data (Yukawa couplings and CKM matrix), we can now proceed to estimate
the contributions to neutral meson mass differences. As stressed
before, we are considering a general multi-Higgs model, so that we
have no definite scheme for the $\mu$-terms and soft SUSY breaking
masses. Therefore, and as done in Section~\ref{higgsphenom}, we will
assume simple textures for the Higgs sector.

\subsection{Tree-level FCNC in neutral mesons: numerical
results}\label{results} 

The first step in evaluating the contributions to tree-level FCNC is
to parameterise the Higgs sector, and thus obtain the mixing matrices
and mass eigenstates entering in
Eqs.~(\ref{MK12:sigma},\ref{MK12:phi}).
As discussed in Section~\ref{higgsphenom}, the Higgs basis is far more
intuitive to work in than the original superpotential basis. Working in
Higgs-basis notation (see Eqs.~(\ref{higgs:Ptransf}-\ref{mij:bij:def})),
let us assume that the free parameters $m^2_{ij}$ and $b_{ij}$ obey
the following simple patterns:
\begin{align}
m^{(d)}_{ij} = \left(
\begin{array}{ccc}
{\otimes}&{\otimes}&{\otimes}\\
{\otimes} & x_3 & y \\
{\otimes} & y & x_{5}
\end{array}\right)\times {1 \text{TeV}}\,,
\quad 
m^{(u)}_{ij} = \left(
\begin{array}{ccc}
{\otimes}&{\otimes}&{\otimes}\\
{\otimes} & x_4 & y \\
{\otimes} & y & x_{6}
\end{array}\right)\times {1 \text{TeV}}\,,
\quad \sqrt{b_{ij}}=\sqrt b \times {1 \text{TeV}}\,,
\end{align}
where the $\otimes$ denotes an entry which is fixed by the
minimisation conditions (cf. Eqs.~(\ref{minima:d},\ref{minima:u})). The
above parametrisation can be further simplified by taking $x_3=x_4$
and $x_6=x_5$.
Therefore, we have the following free parameters associated with the
Higgs sector:
\begin{equation}
x_3\,, \quad x_5\,, \quad \sqrt b\,, \quad  \tan \beta \,,
\end{equation}
and the pattern of VEVs, namely whether they are degenerate or
hierarchical. 
In the numerical examples we will thus consider the
following textures for $m^2_{ij}$ and $b_{ij}$.
\begin{align}
\text{(A)}\,: \quad&\quad 
x_3 = 1\,, \quad x_5 = 5\,, \quad y=0.4\,, \quad \sqrt b =0.3\,,
\nonumber \\
\text{(B)}\,: \quad& \quad
x_3 = 10\,, \quad x_5 = 10\,, \quad y=0.6\,, \quad \sqrt b =0.5\,.
\end{align}
In each case, we will consider several values for $\tan \beta$, taking
it to lie in the range $3 \lesssim \tan \beta \lesssim 12$. 
Moreover, for every $\tan \beta$ we
take to very distinct schemes of VEVs, that while respecting the EW
symmetry breaking conditions and being in agreement with the chosen
$\tan \beta$ value, exhibit either a degenerate pattern 
($w_1=w_3=w_5$, $w_2=w_4=w_6$) or a strongly hierarchical one 
($w_5=10 \,w_3=100 \,w_1$, $w_6=10 \,w_4= 100 \,w_2$).

For the above range of $\tan \beta$, the Higgs spectra (lightest and
heaviest scalar, pseudoscalar and charged states)
is within the following ranges
\footnote{In both cases $m_{1}^s$ and the masses of the heaviest
  states grow monotonically with $\tan \beta$. However, in the case of
  Texture (A), since the mixing induced by $b$ and the diagonal mass
  $x_3$ are of comparable magnitude, as $\tan \beta$ grows, so does
  the mixing between the states, and thus, as pointed out in 
  Section~\ref{higgsspectrum}, the states $h_2^p$ and $h_2^\pm$ can
  become lighter with increasing $\tan \beta$.}:
\begin{align}
&\text{Texture \ (A)}:  \nonumber \\
&79 \, \text{GeV} \lesssim m_{1}^s \lesssim 90 \,\text{GeV}\, ,\quad 
198 \, \text{GeV} \lesssim m_{2}^p \lesssim 409\,\text{GeV}\, ,\quad
417\, \text{GeV} \lesssim m_{2}^\pm \lesssim  203\,\text{GeV}\, ,
\nonumber \\
&\quad\quad \quad \quad\quad \quad 
5009\, \text{GeV} \lesssim m_{6}^s,\ m_{6}^p,\ m_{5}^\pm 
\lesssim 5011\, \text{GeV}\,. \nonumber \\
&\text{Texture \ (B)}:  \nonumber \\
&79 \, \text{GeV} \lesssim m_{1}^s \lesssim 90 \,\text{GeV}\, ,\quad 
908 \, \text{GeV} \lesssim m_{2}^p \lesssim 1547 \,\text{GeV}\, ,\quad
909 \, \text{GeV} \lesssim m_{2}^\pm \lesssim 1549 \,\text{GeV}\, ,
\nonumber \\
&\quad\quad \quad \quad\quad \quad 
10043\, \text{GeV} \lesssim m_{6}^s,\ m_{6}^p,\ m_{5}^\pm 
\lesssim 10047 \text{GeV}\, .
\end{align}

\begin{figure}[t]
 \hspace*{20mm}
     \epsfig{file=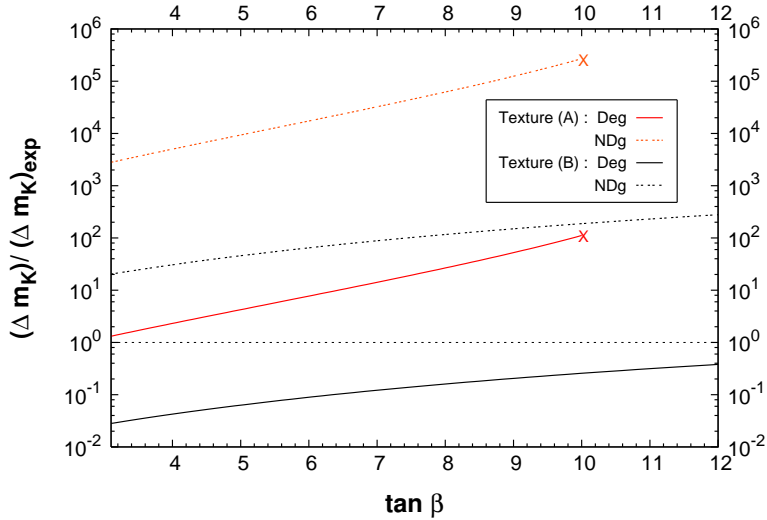,clip=, angle=270,
     width=110mm}
\caption{$\Delta m_K / (\Delta
m_K)_\text{exp}$ as a function of $\tan \beta$. Full (dashed) red
line denotes texture (A) for degenerate (hierarchical) VEVs, while
a black full (dashed) line refers to texture (B) with degenerate
(hierarchical) VEVs. For Texture (A) a cross marks the appearance of
tachyonic states. The experimental bound is depicted via a dotted
horizontal line.}\label{fig:FS.DmK}
\end{figure}

Let us begin by addressing the observable which is associated to most
stringent constraints for this class of models: $\Delta m_K$. In
Fig.~\ref{fig:FS.DmK}, we plot ratio $\Delta m_K / (\Delta
m_K)_\text{exp}$ as a function of $\tan \beta$ for textures (A) and
(B), considering in each case a pattern of degenerate (Deg) or
non-degenerate (NDeg) VEVs.
Fig.~\ref{fig:FS.DmK} clearly reflects the most problematic aspect of
this class of multi-Higgs doublet models. Without a symmetry
forbidding some of the Yukawa couplings, and if the Yukawas themselves
do not exhibit a strong hierarchical character, the contributions to
the neutral kaon mass difference can only be brought down to the
experimental value via a set of very heavy Higgses, as those of
texture (B). A Higgs spectrum closer to the EW scale, with a typical
mass scale of 500 GeV, would generate, for the case of degenerate
(non-degenerate) VEVs, contributions to $\Delta m_K$ of around 
30 ($3 \times 10^{4}$) $(\Delta m_K)_\text{exp}$.  
It is also manifest that smaller values of $\tan \beta$ favour smaller
contributions. This is due to having the Yukawa couplings for the down
quarks proportional to $\sec \beta$.
The relevance of the VEV regime should also be emphasised, 
since the latter plays a very important role. Even though the Yukawas
enter in the contribution to $\mathcal{M}_{12}^K$ already rotated by
the matrices that diagonalise the quark mass matrices, it is clear
that the smaller the VEVs associated to the first and second quark
generations, the more enhanced will be the (12) matrix elements. In
the case of degenerate VEVs, all the Yukawas are identically
suppressed/enhanced\footnote{This is also related to
the specific ansatz for the Yukawa couplings.}.
Naturally, assuming such a large scale for the soft-breaking terms
potentially leads to a fine tuning problem. As pointed out in 
Section~\ref{numv}, and even though the discussion was conducted for a
distinct basis, masses above the few TeV scale are in principle 
within range of a more than 1\% fine tuning. 

When compared to some previous studies, our results are more
severe. However, let us stress that in our analysis we have taken
a few distinct steps. Firstly, and in comparison to the ans\"atze used
in~\cite{Cheng:1987rs}, our Yukawa couplings are quite different,
since accommodating the current CKM matrix data leads to values of
$a_i,\, b_i$ and $c_i$ quite smaller than those previously
considered. This implies that the present Yukawas are less hierarchic.
Moreover, we have explicitly taken into account the values of the
VEVs, and considered the contributions from the exchange of all
physical scalar and pseudoscalar Higgses (not weak interaction
eigenstates), taking into account not only their masses,
but also their mixings. It is also important to mention that the
values of the hadronic matrix elements have been revised in the past
years.

\begin{figure}[t]
 \hspace*{20mm}
     \epsfig{file=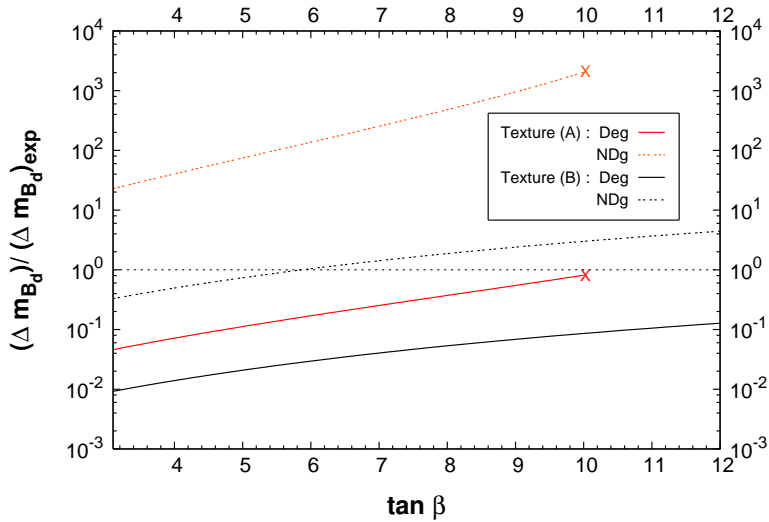,clip=, angle=270, width=110mm}
\caption{$\Delta m_{B_d} / (\Delta
m_{B_d})_\text{exp}$ as a function of $\tan \beta$. Line and colour
codes as in Fig.~\ref{fig:FS.DmK}.}\label{fig:FS.DmB}
\end{figure}

In Fig.~\ref{fig:FS.DmB}, we present the contributions to the mass
difference of the $B_d$ mesons. As expected, in this case it is far
easier to comply with the experimental bounds.
For the case of degenerate VEVs, even the ``lightest'' texture (A)
succeeds in complying with the experimental bounds throughout the
whole range of $\tan \beta$ considered, while for the heavier Higgs
set (B), with non-degenerate VEVs. compatibility is obtained for the
low $\tan \beta$ regime ($\tan \beta \lesssim 6$).

\begin{figure}[t]
 \hspace*{20mm}
     \epsfig{file=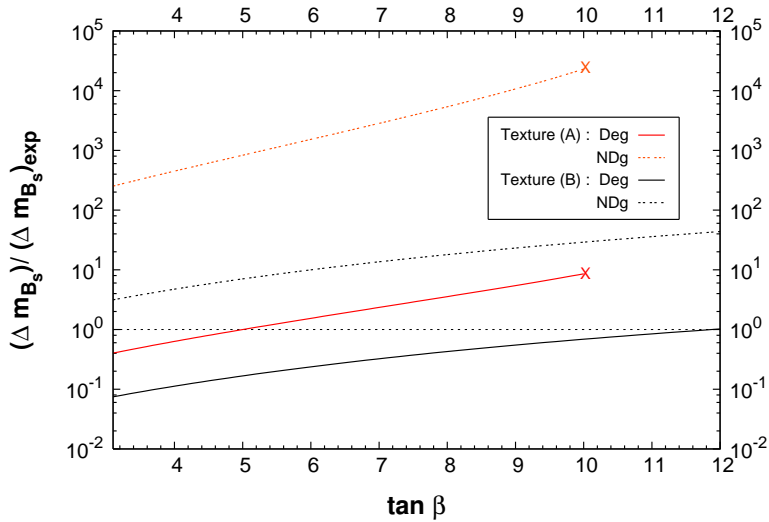,clip=, angle=270, width=110mm}
\caption{$\Delta m_{B_s} / (\Delta
m_{B_s})_\text{exp}$ as a function of $\tan \beta$. Line and colour
codes as in Fig.~\ref{fig:FS.DmK}.}\label{fig:FS.DmBs}
\end{figure}

To complete our study, we display in Fig.~\ref{fig:FS.DmBs} the same
analysis for the case of the $B_s$ meson system. In this case, the
experimental bound is a lower (rather than upper bound). 
As discussed in Section~\ref{fcnc}, the SM already has maximal mixing
in the $B_s$ system, and even though the new contributions may be of
identical magnitude, there is no reason to expect a cancellation
between SM and tree-level contributions, which would be associated to
an extremely fine tuned scenario for all the parameters involved.

From the meson systems so far discussed, the most severe constraints
arise, as expected, from the $K_L - K_S$ mass difference. As a final
remark, and following the
discussion of Section~\ref{fcnc}, we briefly comment on the predicted
scenario involving the $D^0$ mesons. A study similar to those
conducted for the $K^0$ and $B$ sectors is of less interest, since
there is little dependence of $\Delta m_{D} / (\Delta
m_{D})_\text{exp}$ on $\tan \beta$. This is due to having the up-type
quarks Yukawa couplings associated to $\csc \beta$, as seen from 
Eq.~(\ref{fritzsch:def}).

Rather than presenting a plot, we will briefly summarise the
situation.
With the exception of Texture (A), with non-degenerate
VEVs, which induces contributions to $\Delta m_{D}$ around 10 times
its experimental value, all other cases predict contributions below
the experimental bound, ranging from $10^{-3}$ to $10^{-1}$ $(\Delta
m_{D})_\text{exp}$.

To summarise, 
ans\"atze for the Yukawa couplings of the ``simple Fritzsch'' type,
when generalised to multi-Higgs doublet models, typically induce
tree-level FCNC's, and require a Higgs
spectrum at least of order 10 TeV, in order to ensure compatibility with
experiment. 
Notice however that the results presented here are very dependent on
the assumed scheme for the Yukawa couplings, and are only to be taken
as an illustrative example, in the absence of a full theory for the
Higgs-quark-quark interactions. Predictive models for Yukawa
interactions of extended Higgs sector are likely to account for
distinct, and hopefully less severe, bounds~\cite{EMT:quark}.

\section{Conclusions}\label{conc}
Models that predict family replication in the Higgs sector offer a
very aesthetic and phenomenologically interesting scenarios. Although  
there is abundant motivation for extending the Higgs sector (from
string constructions, for example), in most cases the viability of
these models is challenged by the occurrence of potentially dangerous
FCNCs at the tree-level.

We have analysed the most general form of the SUSY potential with
three Higgs families,
studying its minimisation, and deriving the tree-level expressions for
the neutral (scalar and pseudoscalar) and charged Higgses mass matrices.

The main goal of our work was to 
derive a model-independent evaluation of the
tree-level contributions to neutral meson mass differences. We have
computed the most general expression for the tree-level neutral Higgs mediated
contributions to the mass difference of the neutral mesons. We took
into account the exchange of all Higgs states, included the
effects of mixing in the Higgs sector, and made no approximation with
respect to dominant/sub-dominant contributions. 
This analysis is completely general, and can be applied to any given
model with three Higgs families, independently of its Yukawa structure.
As an example, in Section~\ref{flavour:res} we have assumed
a simple ansatz for the Yukawa couplings (in analogy to
what had previously been done), and have considered the contributions of
two distinct Higgs spectra to the $K^0$, $B_d$, $B_s$ and $D^0$ mass
differences, finding that the strongest bound - which as expected
arises from $\Delta m_K$ - requires a spectrum of order 10 TeV.

We again remark that the results for the Higgs masses are strongly
dependent on the specific ansatz for the Yukawa couplings. Other
ans\"atze, that account for a stronger hierarchy in the quark sector
and still accommodate experimental data on quark masses and mixings,
may generate quite smaller contributions to $\Delta m_K$, and thus
require a lighter Higgs spectrum. This possibility is the subject of a
forthcoming work \cite{EMT:quark}, where it is shown that a Higgs
spectra of order 1 TeV can be accommodated. 
On the other hand, it is also
possible that while generating smaller contributions to $\Delta m_K$,
the different ans\"atze induce larger $\Delta m_B$, or $\Delta m_D$.

In addition to the contributions to neutral meson mass difference,
within the quark sector there are other processes that also deserve further
investigation. For example, let us mention, at the one loop
level, the very suppressed SM and MSSM $B_s$
decays. Processes involving the lepton sector also offer an even wider
field for testing the new contributions induced by the additional Higgses
(neutral and charged) \cite{EMT:lepton}. CP violation, given the potential
tree-level contributions to $\varepsilon_K$, may also become a
stringent bound.

\section*{Acknowledgements}
We are grateful to M. Sall\'e for his invaluable help.
The work of N.~Escudero was supported by the ``Consejer\'{\i}a de Educaci\'on
de la Comunidad de Madrid - FPI program'', and ``European Social
Fund''. C.~Mu\~noz acknowledges
the support of the Spanish D.G.I. of the M.E.C. under ``Proyectos
Nacionales'' BFM2003-01266 and FPA2003-04597, and of the European
Union under the RTN program MRTN-CT-2004-503369. The work of A.~M.~Teixeira
is supported by ``Funda\c c\~ao para a Ci\^encia e Tecnologia'' under
the grant SFRH/BPD/11509/2002, and also by  ``Proyectos Nacionales'' 
BFM2003-01266.
The authors are all indebted to KAIST for the hospitality extended to
them during the final stages of this work, and also acknowledge the
financial support of the KAIST Visitor Program.

\end{document}